\documentclass[aps,reprint,longbibliography,superscriptaddress]{revtex4-1}
\usepackage{amsmath}
\usepackage{amssymb}
\usepackage{bm}
\usepackage{epsfig}
\usepackage{graphicx}
\usepackage[dvipsnames]{xcolor}
\usepackage{color}
\usepackage[unicode=true,colorlinks=true,citecolor=blue,urlcolor=blue]{hyperref}
\usepackage[T2A]{fontenc}
\usepackage[cp1251]{inputenc}
\usepackage[english]{babel}
\usepackage{braket}

\newcommand{\erfc}{\mathop{\rm erfc}\nolimits}
\newcommand{\erfcx}{\mathop{\rm erfcx}\nolimits}

\newcommand{\arctanh}{\mathop{\rm arctanh}}
\newcommand{\arctg}{\mathop{\rm arctg}}
\newcommand{\Ci}{\mathop{\rm Ci}}
\newcommand{\Si}{\mathop{\rm Si}}
\newcommand{\Ei}{\mathop{\rm Ei}}

\renewcommand{\i}{{\rm i}}
\renewcommand{\d}{\mathrm d}
\renewcommand{\emph}{\textit}

\renewcommand{\braket}[1]{\left\langle #1 \right\rangle}

\begin{document}

\title{Spin polarization recovery and Hanle effect for charge carriers\\interacting with nuclear spins in semiconductors}

\author{D.~S.~Smirnov}
\email{smirnov@mail.ioffe.ru}
\affiliation{Ioffe Institute, Russian Academy of Sciences, 194021 St. Petersburg, Russia}
\author{E.~A.~Zhukov}
\affiliation{Ioffe Institute, Russian Academy of Sciences, 194021 St. Petersburg, Russia}
\affiliation{Experimentelle Physik 2, Technische Universit\"at Dortmund, 44221 Dortmund, Germany}
\author{D.~R.~Yakovlev}
\affiliation{Experimentelle Physik 2, Technische Universit\"at Dortmund, 44221 Dortmund, Germany}
\affiliation{Ioffe Institute, Russian Academy of Sciences, 194021 St. Petersburg, Russia}
\author{E.~Kirstein}
\affiliation{Experimentelle Physik 2, Technische Universit\"at Dortmund, 44221 Dortmund, Germany}
\author{M.~Bayer}
\affiliation{Experimentelle Physik 2, Technische Universit\"at Dortmund, 44221 Dortmund, Germany}
\affiliation{Ioffe Institute, Russian Academy of Sciences, 194021 St. Petersburg, Russia}
\author{A.~Greilich}
\affiliation{Experimentelle Physik 2, Technische Universit\"at Dortmund, 44221 Dortmund, Germany}

\date{\today}

\begin{abstract}
We report on theoretical and experimental study of the spin polarization recovery and Hanle effect for the charge carriers interacting with the fluctuating nuclear spins in the semiconductor structures. We start the theoretical description from the simplest model of static and isotropic nuclear spin fluctuations. Then we describe the modification of the polarization recovery and Hanle curves due to the anisotropy of the hyperfine interaction, finite nuclear spin correlation time and the strong pulsed spin excitation. For the latter case we describe the resonance spin amplification effect in the Faraday geometry and discuss the manifestations of the quantum Zeno effect. The set of the experimental results for various structures and experimental conditions is chosen to highlight the specific effects predicted theoretically. We show that the spin polarization recovery is a very valuable tool for addressing carrier spin dynamics in semiconductors and their nanostructures.
\end{abstract}

\maketitle

%====================================================

\section{Introduction}

The discovery of the optical orientation in semiconductors by G. Lampel in 1968 set the starting point of the optical spin studies~\cite{Lampel}. Since then, the spin of the charge carriers and related mechanisms of its relaxation, have been a subject of the intense investigations~\cite{OptOR,dyakonov_book}. These are important not only from the fundamental point of view but also for the spintronics applications.

Depending on the specifics of the semiconductor materials and their heterostructures, one can study the carrier spin dynamics by its response to the external influence, such as electric or magnetic fields, temperature, the power and polarization degree of the optical excitation. A variety of methods is available for that: optical orientation~\cite{OptOR}, time-resolved pump-probe schemes to measure a short~\cite{Awschalom_Spintronics,dyakonov_chapter6} or a long time scales~\cite{Extended_pp,Belykh2020}, Hanle effect~\cite{Hanle} and its extension for the pulsed excitation~\cite{Kikkawa98}, the spin inertia method~\cite{PhysRevB.98.125306,PhysRevB.98.121304},  the spin noise spectroscopy~\cite{Zapasskii:13,SinitsynReview}, etc.

Furthermore, intrinsic parameters, like the degree of the carrier localization and the concentration of carriers, determine the degree of influence of concurrent mechanisms on the spin dynamics~\cite{KKavokin-review}. For strongly localized carriers, the significance of the spin-orbit interaction is reduced, while the interaction with the nuclear spin bath becomes decisive~\cite{merkulov02,book_Glazov}.

In this paper, we focus on two methods, where the magnetic field is used to impact the average spin polarization under the depolarizing influence of the unpolarized nuclear spin bath. Spin dynamics in zero or very weak magnetic fields are determined by the hyperfine interaction. By applying the transverse magnetic field (in respect to the spin orientation direction), the Hanle effect is observed. By contrast, the longitudinal magnetic field effectively decouples the carrier spins and the nuclei and results in the increase of the spin polarization, which is described by the so-called polarization recovery curve (PRC). We provide a comparative theoretical and experimental study and analyze a variety of mechanisms controlling the PRC and Hanle signals.

The paper is organized as follows: The theoretical part starts in Sec.~\ref{sec:basic} with discussion of the most basic influence of the nuclear spin fluctuations on the spin dynamics. It is then followed by the discussion of the different extensions. These include the anisotropic hyperfine interaction in Sec.~\ref{sec:Anisotrop} and finite nuclear spin correlation time in Sec.~\ref{sec:Nuclear}. In Sec.~\ref{sec:Pulsed excitation} the additional effects caused by a pulsed excitation are considered.
The experimental part is organized as follows: in Sec.~\ref{subsec:samples} we give a short description of the studied samples; in Sec.~\ref{subsec:techniqes} the experimental techniques used to measure the Hanle and PRC are described; Sec.~\ref{sec:exp_results} presents the experimental results and relates them to the specifics mechanisms. In Sec.~\ref{sec:concl} we conclude the paper.

\section{Basic model}
\label{sec:basic}

\begin{figure}
  \includegraphics[width=\linewidth]{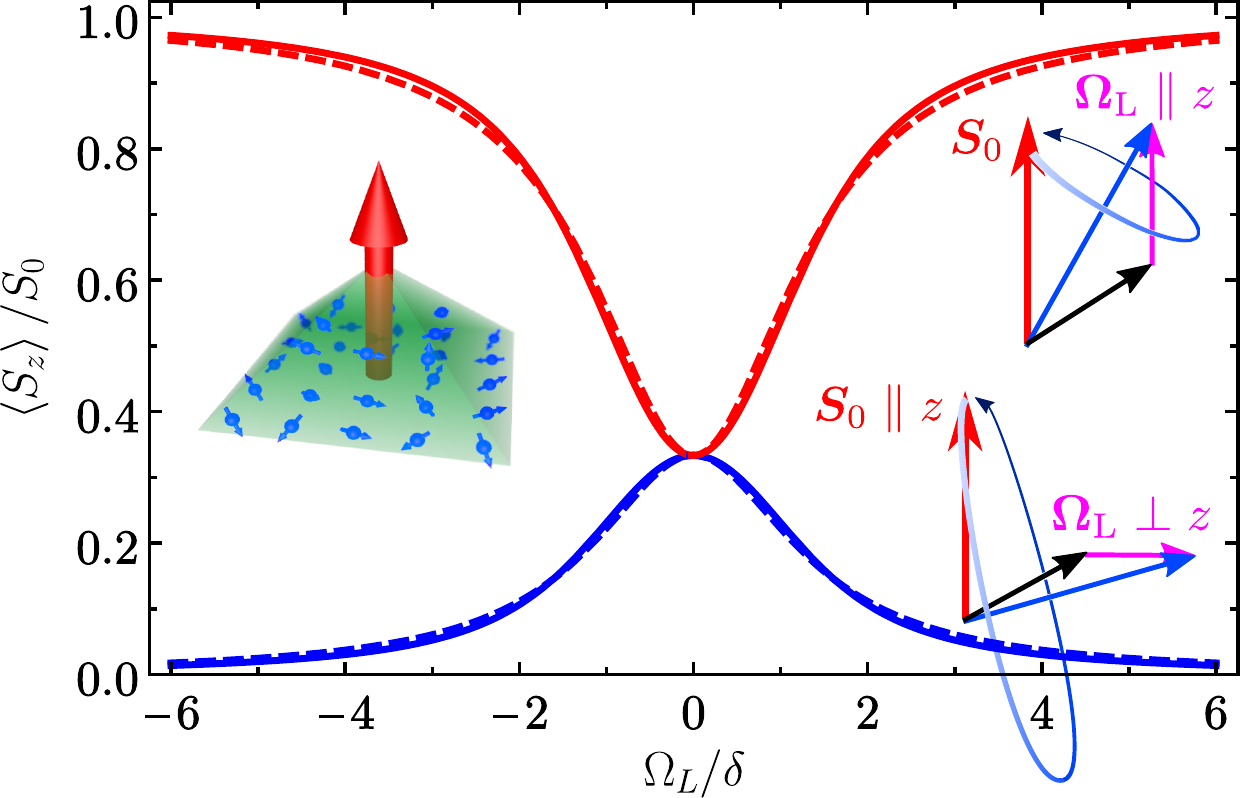}
  \caption{Spin polarization calculated after Eqs.~\eqref{eq:isotropic_H} (solid blue curve) and~\eqref{eq:isotropic_P} (solid red curve) for transverse and longitudinal magnetic field, respectively. The dashed curves show the approximations~\eqref{eq:approx}. The left inset shows the QD with randomly oriented nuclear spins and a single electron spin. The right top and bottom insets illustrate the spin precession with the frequency $\bm\Omega_{\rm tot}$ (blue arrow) in the Faraday and Voigt geometries, respectively.}
  \label{fig:Hanle-PRC}
\end{figure}

Let us consider the basic central spin model (left inset in Fig.~\ref{fig:Hanle-PRC})~\cite{Gaudin}, which captures the essence of the spin polarization recovery and Hanle effects for localized charge carriers~\cite{book_Glazov}. We consider a single localized electron (the case of heavy holes is discussed in Sec.~\ref{sec:Anisotrop}) with the envelop wave function $\Psi(\bm r)$. The present theory also describes homogeneous ensembles of electrons, which are all in the same conditions. The system Hamiltonian includes the hyperfine interaction with the host lattice nuclei and spin interaction with external magnetic field $\bm B$:
\begin{equation}
  \label{eq:H0}
  \mathcal H=\sum_{i} A_i|\Psi(\bm R_i)|^2\upsilon_0\bm I_i\bm S+\hbar\bm\Omega_L\bm S.
\end{equation}
Here $\bm S$ is the electron spin, $i$ enumerates the nuclear spins $\bm I_i$ located at positions $\bm R_i$ with the hyperfine interaction constants $A_i$, $\upsilon_0$ is the unit cell volume, and $\bm\Omega_L=g_e\mu_B\bm B/\hbar$ is the Larmor spin precession frequency of the electron with $g_e$ being the effective electron $g$ factor, $\mu_B$ the Bohr magneton, and $\hbar$ the reduced Planck constant. In this Hamiltonian we assume the hyperfine interaction and the electron $g$ factor to be isotropic, which is usually the case for the electrons in GaAs-like semiconductors. We neglect the nuclear Zeeman splitting, because it is much smaller than the electron one (see Sec.~\ref{sec:Nuclear} for its possible effects). The Hamiltonian~\eqref{eq:H0} can be rewritten as
\begin{equation}
  \mathcal H=\hbar(\bm\Omega_N+\bm\Omega_L)\bm S,
\end{equation}
where
\begin{equation}
  \label{eq:Omega_N}
  \bm\Omega_N=\frac{1}{\hbar}\sum_{i} A_i|\Psi(\bm R_i)|^2\upsilon_0\bm I_i
\end{equation}
is the resident charge carrier spin precession frequency in the Overhauser field. We recall that here we consider the localized electrons, while the holes will be considered in the next section.
% electron spin precession frequency in the Overhauser field $\bm B_N=\hbar\bm\Omega_N/(g_e\mu_B)$. %Sometimes $\bm\Omega_N$ is called nuclear field as well for brevity~\cite{NoiseGlazov}.

In the absence of nuclear spin polarization, the nuclear field is zero on average. However, due to the finite number of nuclei interacting with the localized electron, there are stochastic nuclear spin fluctuations, which are characterized by the probability distribution function
\begin{equation}
  \label{eq:F}
  \mathcal F(\bm\Omega_N)=\frac{1}{(\sqrt{\pi}\delta)^3}\exp\left(-\frac{\Omega_N^2}{\delta^2}\right),
\end{equation}
where $\delta$ determines the dispersion: $\braket{\Omega_N^2}=3\delta^2/2$, with the angular brackets denoting the statistical averaging. In the theoretical analysis we use the angular frequencies only. For the independent and randomly oriented nuclear spins from Eq.~\eqref{eq:Omega_N} we obtain the typical electron spin precession frequency in the nuclear field
\begin{equation}
  \delta=\frac{\upsilon_0}{\hbar}\sqrt{\frac{2}{3}\sum_i A_i^2|\Psi(\bm R_i)|^4I_i(I_i+1)}.
\end{equation}

Usually, the number of nuclei in the electron localization volume is large. Each nuclear spin precesses with the frequency $A_i|\Psi(\bm R_i)|^2\upsilon_0/\hbar$, which is much smaller than $\delta$. Therefore, the nuclear spins can be considered as ``frozen'' on the time scale of the electron spin precession~\cite{Werner1977,merkulov02}. As a result, the electron spin dynamics can be described as precession with the constant frequency $\bm\Omega_{\rm tot}=\bm\Omega_N+\bm\Omega_L$. The spin dynamics $\bm S(t)$ should be averaged over the distribution function~\eqref{eq:F} to obtain the average signal for many repeated measurements or many localized electrons~\cite{merkulov02}.

We consider the spin polarization $S_0$ created at $t=0$ along the $z$ axis. After spin initialization, the spin precession begins. Due to the random magnitude of the nuclear field, the electron spin component perpendicular to $\bm\Omega_{\rm tot}$ dephases during the time $T_2^*\sim1/\delta$. At longer times, the average spin is contributed only by the spin component parallel to $\bm\Omega_{\rm tot}$:
\begin{equation}
  \label{eq:Sz_average}
  \braket{S_z}=\braket{\frac{\Omega_{{\rm tot},z}^2}{\Omega_{\rm tot}^2}}S_0.
\end{equation}
The two other spin components are zero on average: $\braket{S_x}=\braket{S_y}=0$. In the model under consideration the average spin in Eq.~\eqref{eq:Sz_average} does not decay with time. In every real system, additional mechanisms unrelated with the hyperfine interaction, such as spin-orbit and electron-phonon interactions~\cite{PhysRevLett.88.186802,PhysRevB.94.125401}, destroy it during a spin relaxation time, which we denote $\tau_s$. It will be important for the results described in Sec.~\ref{sec:Nuclear}, and will be discussed there in more detail. Here we assume that $\tau_s\gg1/\delta$, so Eq.~\eqref{eq:Sz_average} holds at $1/\delta\ll t\ll\tau_s$.

The Hanle and PRC are given by the dependence of the average spin $\braket{S_z}$ on the magnetic field in Voigt and Faraday geometries, respectively (the light propagation axis coincides with the direction of the spin polarization $z$). For convenience, we denote the ratio $\braket{S_z}/S_0$ as $H(\Omega_L)$ and $P(\Omega_L)$ for these two cases, respectively. Calculation of the average in Eq.~\eqref{eq:Sz_average} yields
\begin{subequations}
  \label{eq:isotropic}
  \begin{equation}
    \label{eq:isotropic_H}
    H(\Omega_L)=\frac{\delta^2}{2\Omega_L^2}\left[1-\frac{\delta}{\Omega_L}D\left(\frac{\Omega_L}{\delta}\right)\right],
  \end{equation}
  \begin{equation}
    \label{eq:isotropic_P}
    P(\Omega_L)=1-2H(\Omega_L),
  \end{equation}
\end{subequations}
where $D(x)=\exp(-x^2)\int_0^x \exp(y^2)\d y$ is the Dawson integral. Noteworthy, in this model, $P(\Omega_L)+2H(\Omega_L)=1$. The corresponding curves are shown by solid lines in Fig.~\ref{fig:Hanle-PRC}. One can see that for zero magnetic field
\begin{equation}
  \label{eq:1/3}
  \braket{S_z}=S_0/3.
\end{equation}

In the limit of the strong transverse magnetic field one has for the Hanle curve
\begin{equation}
  \braket{S_z}=0.
\end{equation}
Qualitatively, in the strong transverse magnetic field, $\Omega_L\gg\delta$, the total spin precession frequency $\bm\Omega_{\rm tot}$ is parallel to the magnetic field, so its $z$ component vanishes, see upper right inset in Fig.~\ref{fig:Hanle-PRC}. In this case the initial spin polarization $S_0$ dephases completely, and Eq.~\eqref{eq:isotropic_H} yields zero.

In the limit of a strong longitudinal magnetic field, the total spin precession frequency is parallel to the $z$ axis, lower right inset in Fig.~\ref{fig:Hanle-PRC}. So, for $\Omega_L \gg \delta$, one has
\begin{equation}
  \label{eq:Sz0}
  \braket{S_z}=S_0.
\end{equation}
In this case, the initial spin polarization does not dephase and Eq.~\eqref{eq:isotropic_P} yields Eq.~\eqref{eq:Sz0}.

In zero magnetic field, one can say that the nuclear field can be either parallel to $x$, $y$, or $z$ axis. In the first two cases the spin polarization dephases completely, while for the latter one it does not dephase. As a result one obtains 1/3 of the initial spin polarization, Eq.~\eqref{eq:1/3}.

The Hanle and PRC given by Eqs.~\eqref{eq:isotropic} can be approximated by the Lorentzians:
\begin{subequations}
  \label{eq:approx}
  \begin{equation}
    \label{eq:H_main}
  H(\Omega_L)\approx\frac{2}{3}\frac{\delta^2}{2\delta^2+\Omega_L^2},
  \end{equation}
  \begin{equation}
    \label{eq:P_main}
  P(\Omega_L)\approx\frac{1}{3}\frac{2\delta^2+3\Omega_L^2}{2\delta^2+\Omega_L^2}.
 \end{equation}
\end{subequations}
These approximations are shown in Fig.~\ref{fig:Hanle-PRC} by the dashed curves and agree very well with the exact calculations (the maximum difference is of the order of $1$\%).

In this section, we provided the basic description of the Hanle and polarization recovery effects. In the following sections, we introduce various generalizations of this model.

\section{Anisotropic hyperfine interaction}
\label{sec:Anisotrop}

In Sec.~\ref{sec:basic} we assumed that the distribution function of the spin precession frequency $\bm\Omega_N$ is isotropic, Eq.~\eqref{eq:F}. However, for example, in GaAs type semiconductors for holes in the $\Gamma$ valley~\cite{book_Glazov} and for electrons in the X valley~\cite{ShchepetilnikovNMR,PhysRevB.101.075412} the hyperfine interaction is anisotropic. Generally, the random nuclear field is described by the distribution function
\begin{equation}
  \label{eq:F_lambda}
  \mathcal F(\bm\Omega_N)=\frac{1}{\pi^{3/2}\delta_x\delta_y\delta_z}\exp\left(-\frac{\Omega_{N,x}^2}{\delta_x^2}-\frac{\Omega_{N,y}^2}{\delta_y^2}-\frac{\Omega_{N,z}^2}{\delta_z^2}\right),
\end{equation}
where $\delta_{x}$, $\delta_{y}$  and $\delta_{z}$  are independent parameters.

%Here we describe the Hanle and polarization recovery effects for the case, when two of the three parameters coincide.

\begin{figure}
  \includegraphics[width=\linewidth]{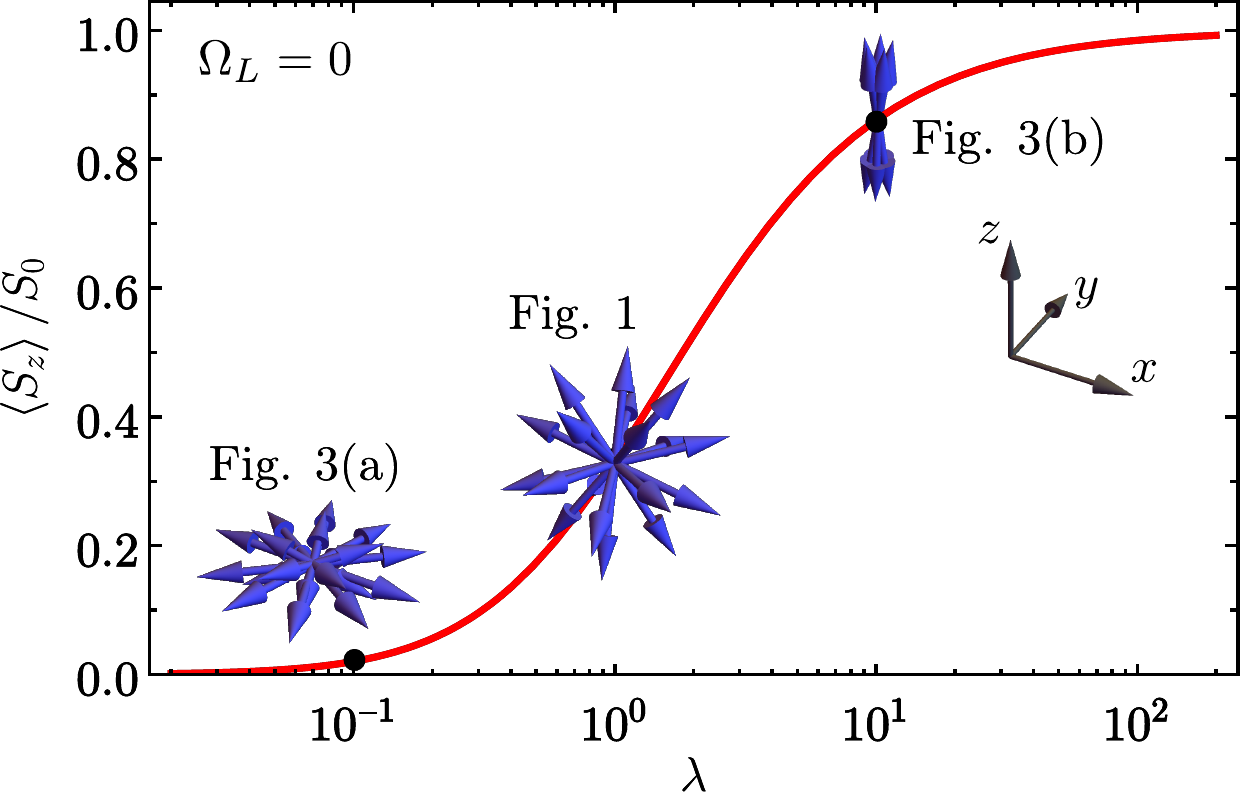}
  \caption{The spin polarization in zero magnetic field, $\Omega_L=0$, calculated after Eq.~\eqref{eq:f} as a function of the anisotropy parameter $\lambda$. The schematics show the distributions of the random nuclear field for the corresponding values of $\lambda$ in the coordinate frame shown in the inset, see Eq.~\eqref{eq:F_lambda}.}
  \label{fig:ani}
\end{figure}

Let us consider $\delta_x=\delta_y=\delta$ and $\delta_z=\lambda\delta$ with $\lambda$ being the anisotropy parameter, which is relevant for heavy holes.
%In particular, for $\lambda>1$ in the absence of magnetic field the spin polarization is larger than $S_0/3$.
In zero magnetic field, from Eqs.~\eqref{eq:Sz_average} and~\eqref{eq:F_lambda} we obtain the dependence of the spin polarization $\braket{S_z}$ on the anisotropy parameter $\lambda$~\cite{MX2_Avdeev}
\begin{equation}
  \label{eq:f}
  \frac{\braket{S_z}}{S_0}=\left\lbrace
    \begin{array}{lr}
      \dfrac{\lambda^2\left(\sqrt{\lambda^2-1}-\arctan\sqrt{\lambda^2-1}\right)}{\left(\lambda^2-1\right)^{3/2}}, & \lambda>1\\
      1/3, & \lambda=1\\
      \dfrac{\lambda^2\left(-\sqrt{1-\lambda^2}+\arctanh\sqrt{1-\lambda^2}\right)}{\left(1-\lambda^2\right)^{3/2}}, & \lambda<1
    \end{array}
  \right.
  \:,
\end{equation}
the cases of $\lambda<1$ and $\lambda>1$ can be obtained one from another by the analytic continuation. Note the difference between $\arctan(x)$ and $\arctanh(x)$ functions. This dependence is shown in Fig.~\ref{fig:ani}. The spin polarization monotonously increases from $0$ to $1$ as $\lambda$ varies from $0$ to~$\infty$.

\begin{figure}
  \includegraphics[width=\linewidth]{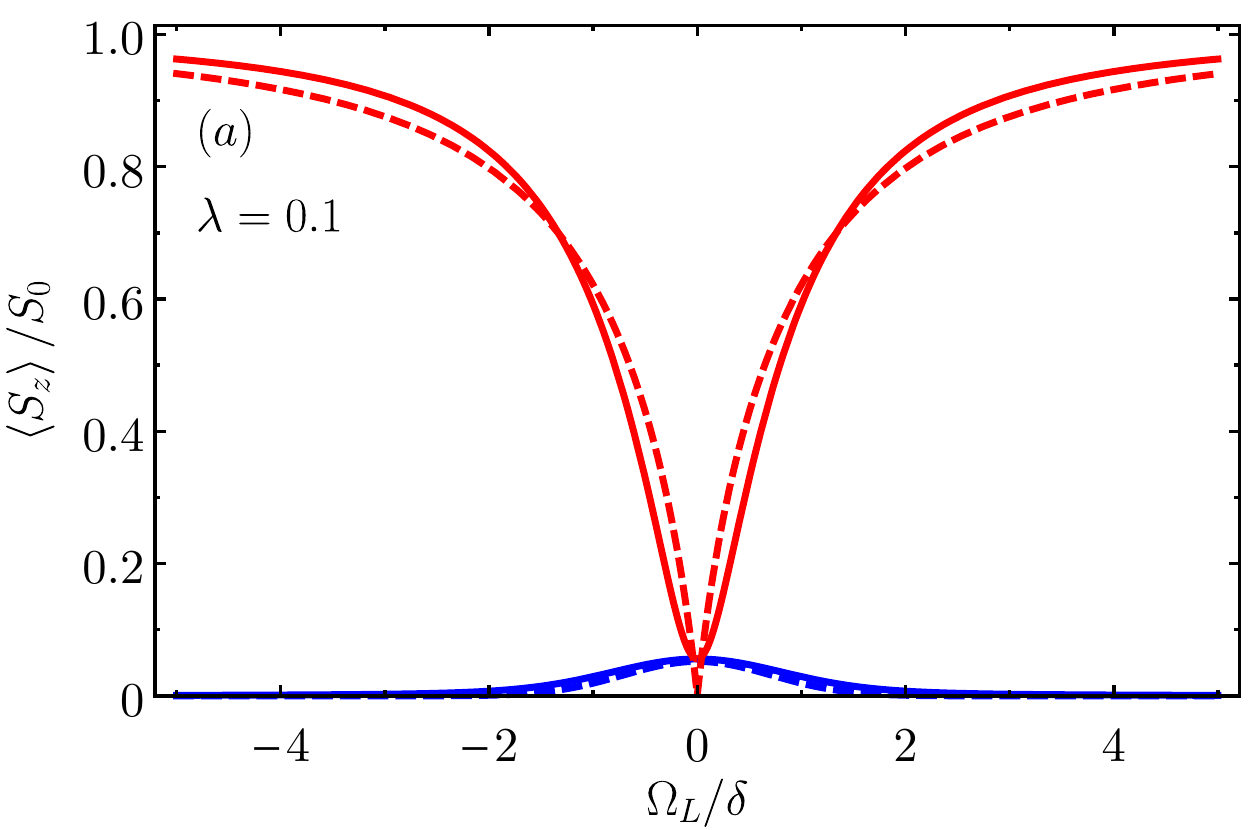}\\[2mm]
  \includegraphics[width=\linewidth]{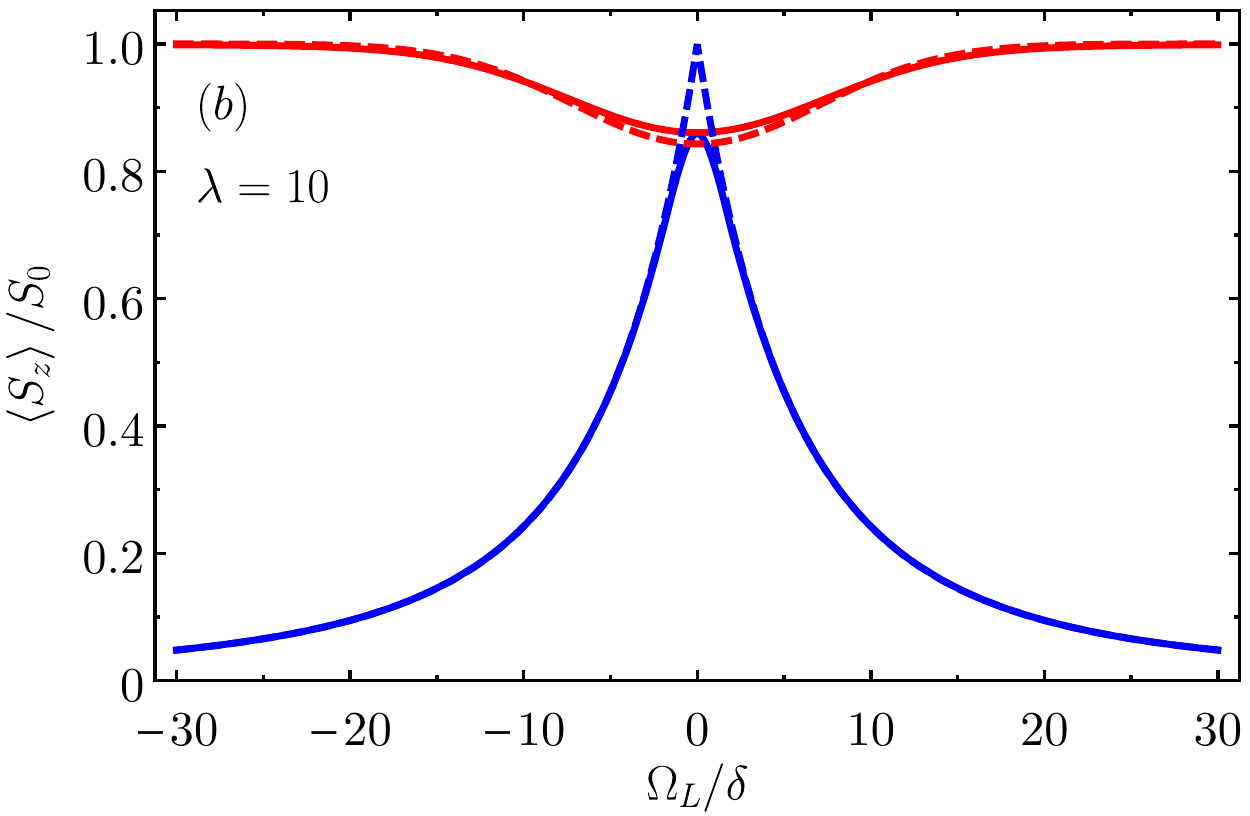}
  \caption{(a) Hanle) and PRC (red) calculated after Eq.~\eqref{eq:Sz_average} for the parameters $\delta_x=\delta_y=\delta$ and $\delta_z=0.1\delta$ ($\lambda=0.1$). The dashed curves show Eqs.~\eqref{eq:z_weak}. (b) The same curves for $\delta_x=\delta_y=\delta$ and $\delta_z=10\delta$ ($\lambda=10$) and approximations~\eqref{eq:z_strong}.}
  \label{fig:z}
\end{figure}

In the limit $\lambda\ll 1$, the nuclear field is distributed in the $(xy)$ plane (see the corresponding inset in Fig.~\ref{fig:ani}), so the average spin polarization is zero, similarly to the case of strong transverse magnetic field. In the limit $\lambda\gg 1$, the nuclear field is parallel to the $z$ axis (see the corresponding inset in Fig.~\ref{fig:ani}), so $\braket{S_z}=S_0$, similarly to the strong longitudinal magnetic field. In the case of $\lambda=1$ one has $\braket{S_z}=S_0/3$, in agreement with Eq.~\eqref{eq:1/3} for the isotropic case.

The Hanle and PRC can be described analytically in the case of the strong anisotropy. In the limit $\lambda\ll1$, we obtain
\begin{subequations}
  \label{eq:z_weak}
  \begin{equation}
    H(\Omega_L)=\lambda^2\left[\ln(2/\lambda)-1\right]\exp\left(-\frac{\Omega_L^2}{\delta_z^2}\right)\ll 1,
  \end{equation}
  \begin{multline}
    P(\Omega_L)=\frac{|\Omega_L|}{\delta}\left\lbrace\sin\left(\frac{|\Omega_L|}{\delta}\right)\Ci\left(\frac{|\Omega_L|}{\delta}\right)
      \right.\\\left.
        +\cos\left(\frac{|\Omega_L|}{\delta}\right)\left[\frac{\pi}{2}-\Si\left(\frac{|\Omega_L|}{\delta}\right)\right]\right\rbrace,
  \end{multline}
\end{subequations}
where $\Si(x)$ and $\Ci(x)$ are sine and cosine integral functions, respectively. These expressions are shown in Fig.~\ref{fig:z}(a) by blue and red dashed curves. One can see that in this limit the spin polarization in zero magnetic field is small: $\braket{S_z}/S_0\ll 1$, as discussed above. The typical width of the both curves is of the order of $\delta$. In this limit the Hanle curve is Gaussian, while the PRC is similar to the sharp Lorentzian form. The solid curves in Fig.~\ref{fig:z}(a) show the numerical calculations after Eqs.~\eqref{eq:Sz_average} and~\eqref{eq:F_lambda} for $\lambda=0.1$ for comparison.

In the opposite limit of $\lambda\gg1$, we find the approximate expressions
\begin{subequations}
  \label{eq:z_strong}
  \begin{equation}
    H(\Omega_L)=1-\sqrt{\pi}\frac{|\Omega_L|}{\delta_z}\exp\left(\frac{\Omega_L^2}{\delta_z^2}\right)\erfc\left(\frac{|\Omega_L|}{\delta_z}\right),
  \end{equation}
  \begin{equation}
    P(\Omega_L)=1-\frac{\pi}{2\lambda}\exp\left(-\frac{\Omega_L^2}{\delta_z^2}\right)\approx 1.
  \end{equation}
\end{subequations}
These results are shown in Fig.~\ref{fig:z}(b). The typical width of the curves is of the order of $\delta_z=\lambda\delta\gg\delta$ [note the difference in the scales of the horizontal axes in panels (a) and (b)]. The shapes of the curves are opposite to the previous limit: PRC is Gaussian, and Hanle curve is sharp, as one can see from Eqs.~\eqref{eq:z_strong}.

Generally, one can say that the polarization recovery and suppression require $\Omega_L$ to be larger than the largest parameter among $\delta_x$, $\delta_y$, and $\delta_z$.

One can describe in a similar way the cases of $\delta_x\neq\delta_y=\delta_z$ or $\delta_y\neq\delta_x=\delta_z$. This situation is relevant for electrons in the $X$ valley. In (In,Al)As/AlAs quantum dots (QDs), the lowest state of the conduction band can belong to one of the two $X$ valleys oriented along $[100]$ and $[010]$ crystallographic axes~\cite{Shamirzaev78}. In this case, the hyperfine interaction is stronger along the valley axis than along the two other directions~\cite{PhysRevB.101.075412}. The Hanle curve is anisotropic and its width depends on the relative orientation of the magnetic field and the valley direction. If the two in-plane $X$ valleys are equally populated, this anisotropy can be hidden in experiment. But strain applied along $[100]$ or $[010]$ axis can lead to the valley splitting and preferential occupation for one of them, which will uncover the hidden anisotropy.

Experimentally, electrons and holes can coexist in the same sample~\cite{Glasenapp2016}. Provided they are independent, their contributions to the Hanle and PRC should be summed up. It is important to discuss the relative signs of these two contributions to the observed spin signals. We assume that the spin polarization is created resonantly using the optical orientation~\cite{ivchenko05a}. Afterwards the spin polarization can be probed via ellipticity of Faraday rotation of linearly polarized probe pulses. Due to the optical selection rules, only one of the electron and hole spin components interacts with the light of the given helicity. This determines the signs of spin orientation and the ellipticity signal both for electrons and holes~\cite{yugova09}. As a results, the ellipticity has always the same sign for electrons and holes. The sense of the Faraday rotation is determined by the detuning of the probe pulses from the optical resonance of the QDs, and it is the same for the same sign of the detuning. Due to the inhomogeneous broadening in the ensembles of localized electrons and holes, they can provide the Faraday rotation angle both of the same and opposite signs.

To summarize this section, the anisotropy of the hyperfine interaction can lead to the suppression of either Hanle or polarization recovery effect and to the change of the shapes of these dependencies. The widths of these curves are of the same order and are determined by the largest component of the nuclear field.

\section{Nuclear spin correlation time}
\label{sec:Nuclear}

In the previous sections, we limited ourselves to the model of frozen nuclear spin fluctuations. Generally, the spin of an electron obeys the Bloch equation
\begin{equation}
  \label{eq:Bloch}
  \frac{\d\bm S}{\d t}=\left[\bm\Omega_N(t)+\bm\Omega_L\right]\times\bm S-\frac{\bm S}{\tau_s},
\end{equation}
where $\tau_s$ is the spin relaxation time introduced in Sec.~\ref{sec:basic}. Eq.~\eqref{eq:Bloch} implies that the electron longitudinal ($T_1$) and transverse ($T_2$) spin relaxation times are equal. The average spin polarization is given by
\begin{equation}
  \label{eq:average_spin}
  \braket{S_z}\equiv\int\limits_0^\infty S_z(t)\frac{\d t}{\tau_s}.
\end{equation}
If the time dependence of $\bm\Omega_N(t)$ can be neglected and $\tau_s\delta\gg 1$, this definition coincides with the average used in the previous sections, and the corresponding results are valid. If by contrast the spin relaxation time $\tau_s$ is much shorter than $1/\delta$, there is no polarization recovery effect, and the Hanle curve $H(\Omega_L)$ has a simple Lorentzian form with the width $1/\tau_s$.

In this section we will study the role of the finite nuclear spin correlation time $\tau_c$ under the assumption of $\tau_s\delta\gg 1$ (for the isotropic hyperfine interaction).

Typically, $\tau_c$ is also longer than the typical electron spin precession period ($1/\delta$). The nuclear spin dynamics can be caused by the Knight field of the electrons, by the nuclear dipole-dipole interaction, or by the interaction of the nuclear quadrupole moment with the strain and random electric fields in the structure~\cite{book_Glazov,Dzhioev2007,PhysRevB.94.081201}.

We consider the simplest model of nuclear spin dynamics, which assumes random abrupt changes of the nuclear field with the typical correlation time $\tau_c$ between the states described by the distribution function~\eqref{eq:F}~\cite{PhysRevB.98.125306}. In this case the noise of $\bm\Omega_N(t)$ is the telegraph noise.

Solution of Eq.~\eqref{eq:Bloch} averaged over the nuclear fields can be found using the Fourier transform~\cite{schulten,Glazov_hopping}. The average spin defined by Eq.~\eqref{eq:average_spin} is given by the Fourier component at zero frequency. The result for the both transverse and longitudinal magnetic fields can be written in the form:
\begin{equation}
  \label{eq:HPtauc}
  \frac{\braket{S_z}}{S_0}=\frac{(1-\mathcal A)(\mathcal A+\tau_c/\tau_s)\tau_c/\tau_s-\mathcal B^2/(1+\tau_s/\tau_c)}{(\mathcal A+\tau_c/\tau_s)^2+\mathcal B^2},
\end{equation}
where
\begin{subequations}
  \begin{equation}
    \mathcal A=\braket{\frac{\Omega_{{\rm tot},x}^2+\Omega_{{\rm tot},y}^2}{\Omega_{\rm tot}^2+(1/\tau_s+1/\tau_c)^2}},
  \end{equation}
  \begin{equation}
    \mathcal B=\frac{1}{\Omega_L}\left(\frac{1}{\tau_c}+\frac{1}{\tau_s}\right)\braket{\frac{\bm\Omega_L\bm\Omega_{\rm tot}}{\Omega_{\rm tot}^2+(1/\tau_s+1/\tau_c)^2}}
  \end{equation}
\end{subequations}
with angular brackets denoting average over the distribution function~\eqref{eq:F}, as above.

For the longitudinal magnetic field we obtain $\mathcal B=0$ and
\begin{equation}
  \mathcal A=\frac{\delta^2}{2\Omega_L^2}+\frac{\sqrt{\pi}\delta}{4\Omega_L^2}\erfcx(\xi)\left(\i\frac{\delta^2}{\Omega_L}-\frac{2}{\tau_c}-\frac{2}{\tau_s}\right)+\rm{c.c.},
\end{equation}
where $\xi=(1/\tau_c+1/\tau_s)/\delta-\i\Omega_L/\delta$ and $\erfcx(\xi)=\exp(\xi^2)\erfc(\xi)$ is the scaled complementary error function. Similarly, for transverse magnetic field we obtain
\begin{subequations}
  \begin{multline}
    \mathcal A=\frac{1}{2}-\frac{\delta^2}{4\Omega_L^2}+\frac{\sqrt{\pi}\delta^3}{8\Omega_L^3}\erfcx(\xi)\\
    \times\left[-\i+2\frac{\Omega_L}{\delta^2}\left(\frac{1}{\tau_c}+\frac{1}{\tau_s}\right)+4\i\left(\frac{\Omega_L}{\delta^2}\right)^2\left(\frac{1}{\tau_c}+\frac{1}{\tau_s}\right)^2\right]+\rm{c.c.},
  \end{multline}
  \begin{equation}
    \mathcal B=\frac{\sqrt{\pi}\i}{2\Omega_L}\left(\frac{1}{\tau_c}+\frac{1}{\tau_s}\right)\erfcx(\xi)+\rm{c.c.}.
  \end{equation}
\end{subequations}
Thus the Hanle and polarization recovery effects for the finite nuclear spin correlation time can be described analytically, although the expressions are cumbersome.

In the limit $\tau_c\gg\tau_s$ the nuclear spin dynamics can be neglected, so Eq.~\eqref{eq:HPtauc} reduces to Eqs.~\eqref{eq:isotropic}.

\begin{figure}
 \includegraphics[width=\linewidth]{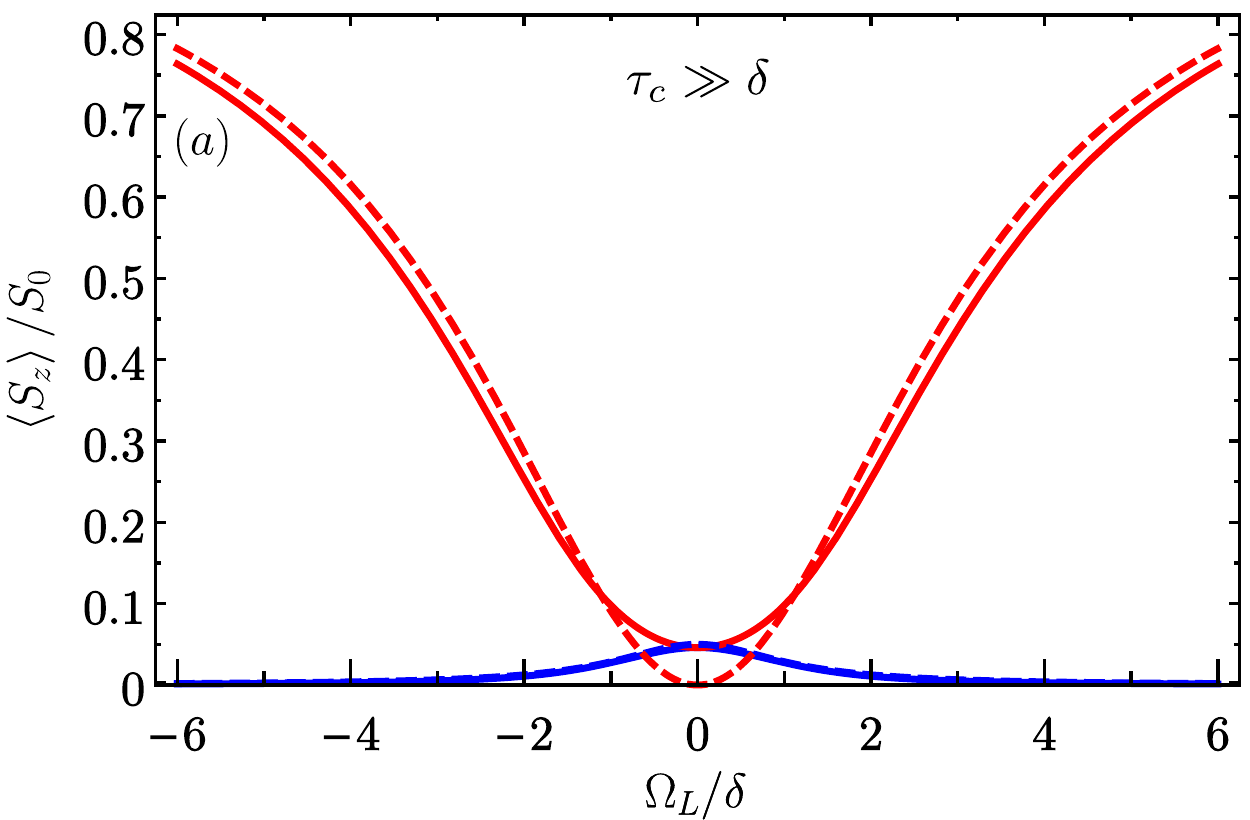}\\[2mm]
  \includegraphics[width=\linewidth]{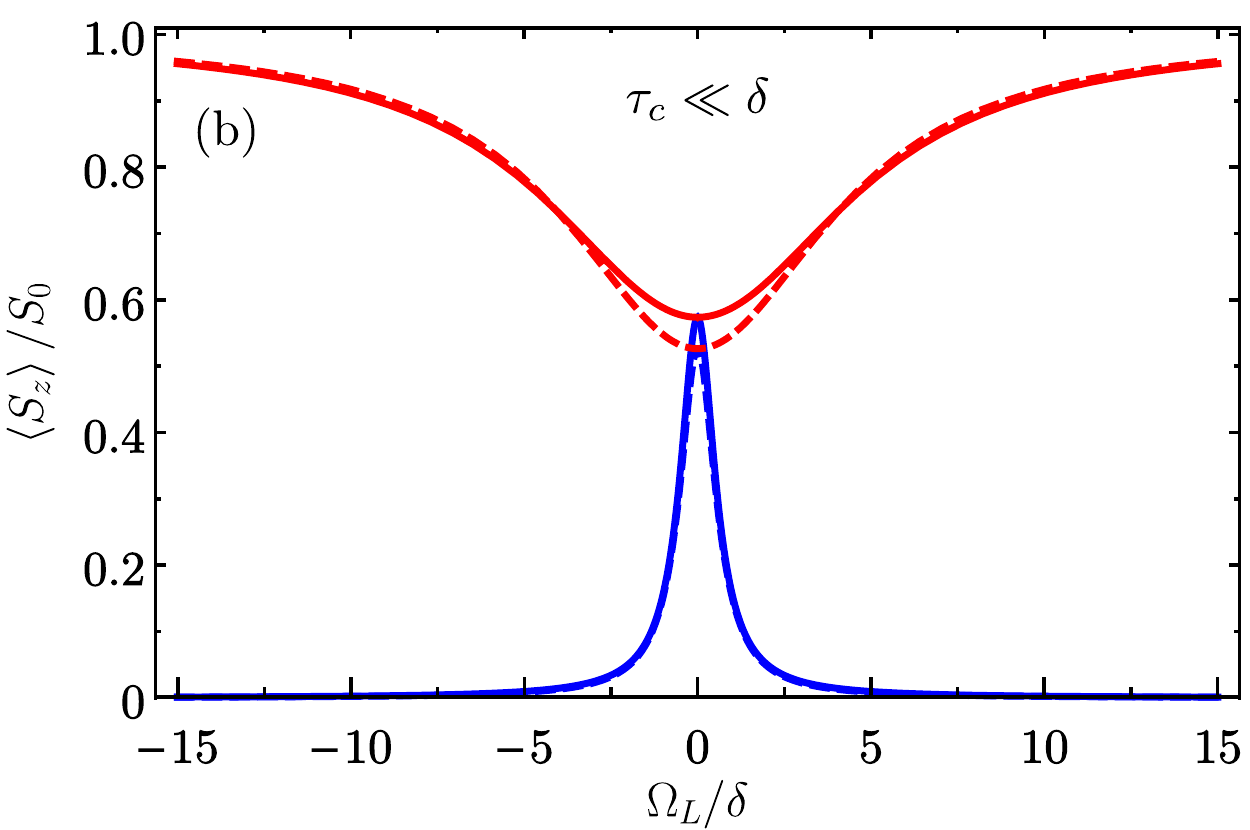}
  \caption{Hanle (blue) and PRC (red) for finite nuclear spin correlation time calculated after Eq.~\eqref{eq:HPtauc} for the parameters (a) $\tau_s\delta=100$, $\tau_c\delta=10$; (b) $\tau_s\delta=3$, $\tau_c\delta=0.3$. The dashed curves show Eqs.~\eqref{eq:Htauc_int2} and~\eqref{eq:Ptauc_int} in panel (a) and~\eqref{eq:short_tau} in panel (b).}
  \label{fig:tau_c}
\end{figure}

For the moderate nuclear spin correlation time, $\tau_s\gg\tau_c\gg1/\delta$, for zero magnetic field we find from Eq.~\eqref{eq:average_spin}
\begin{equation}
  \frac{\braket{S_z}}{S_0}=\frac{\tau_c}{2\tau_s},
\end{equation}
which is small. Indeed, in this regime the spin polarization disappears on average after a few nuclear field reorientations at $t\sim\tau_c$, which is much shorter than $\tau_s$. For transverse and longitudinal magnetic fields we obtain
%\begin{subequations}
  \begin{equation}
    \label{eq:Htauc_int}
    H(\Omega_L)=\frac{\tau_c}{\tau_s}\frac{\Omega_L/\delta-D(\Omega_L/\delta)}{2(\Omega_L/\delta)^3-\Omega_L/\delta+D(\Omega_L/\delta)},
    % H(\Omega_L)=\frac{\tau_c}{\tau_s}\frac{\braket{\Omega_{{\rm tot},z}^2/\Omega_{\rm tot}^2}}{\braket{(\Omega_{{\rm tot},x}^2+\Omega_{{\rm tot},y}^2)/\Omega_{\rm tot}^2}},
  \end{equation}
  \begin{equation}
    \label{eq:Ptauc_int}
    P(\Omega_L)=\frac{\Omega_L^2}{\Omega_L^2+\delta^2\tau_s/\tau_c},
  \end{equation}
where the Dawson function $D(x)$ is defined below Eq.~\eqref{eq:isotropic}. Using the numeric approximations~\eqref{eq:approx}, one can also rewrite the former expression as follows:
\begin{equation}
  \label{eq:Htauc_int2}
  H(\Omega_L)\approx\frac{\tau_c}{\tau_s}\frac{2\delta^2}{4\delta^2+3\Omega_L^2}\ll 1.
%  \tag{\ref*{eq:Htauc_int}${}^\prime$}
  % \tag{\ref*{eq:Htauc}'}
\end{equation}
%\end{subequations}
From these equations one can see that Hanle and PRC are Lorentzian in this limit. The comparison between exact and approximate expressions is shown in Fig.~\ref{fig:tau_c}(a).

Finally, the limit of short nuclear spin correlation time, $\tau_c\ll1/\delta$, is similar to the Dyakonov-Perel spin relaxation~\cite{dyakonov_book}, because the electron spin precession frequency changes faster than the typical precession period. The spin dynamics in this regime is well known~\cite{dyakonov_book}. It represents the monoexponential relaxation with the rate
\begin{equation}
  \label{eq:tau_s_eff}
  \frac{1}{\tau_{\rm eff}}=\frac{1}{\tau_s}+\delta^2\tau_c
\end{equation}
for zero magnetic field. From Eq.~\eqref{eq:HPtauc} we find the Hanle curve and PRC
\begin{subequations}
  \label{eq:short_tau}
  \begin{equation}
    H(\Omega_L)=\frac{\tau_{\rm eff}/\tau_s}{1+(\Omega_L\tau_{\rm eff})^2},
  \end{equation}
  \begin{equation}
    P(\Omega_L)=\frac{1+(\Omega_L\tau_c)^2}{\tau_s/\tau_{\rm eff}+(\Omega_L\tau_c)^2}.
  \end{equation}
\end{subequations}
These expressions and a comparison with the numeric calculations are shown in Fig.~\ref{fig:tau_c}(b). The Hanle and PRC in this limit are Lorentzian again. At zero magnetic field from Eqs.~\eqref{eq:short_tau} we obtain
\begin{equation}
  \frac{\braket{S_z}}{S_0}=\frac{\tau_{\rm eff}}{\tau_s},
\end{equation}
which can be in the range from $0$ to $1$, depending on the dominant spin relaxation mechanism in Eq.~\eqref{eq:tau_s_eff}. Curiously, the widths of the two curves are parametrically different in this case. The Hanle curve is narrow with the half width at half maximum (HWHM) $1/\tau_{\rm eff}$ as in the Hanle effect without hyperfine interaction. In the same time, the PRC is wide with the HWHM $1/\tau_c\gg1/\tau_{\rm eff}$. This is because the spin precession frequency in the longitudinal magnetic field must be larger than the nuclear spin correlation time to suppress nuclei induced spin relaxation of electrons~\cite{dyakonov_book}.

To summarize this section, the nuclear spin dynamics can increase or decrease the spin polarization in zero magnetic field (relative to $S_0/3$) and it makes the PRC broader than the Hanle curve.

%\commentDima{This section can be illustrated by the experiments with $n$-type QDs.}
%\commentDima{Experiments from Secs.~\ref{sec:exp_4}, \ref{sec:exp_6} and~\ref{sec:exp_9}.}

\section{Pulsed spin excitation}
\label{sec:Pulsed excitation}

In the previous sections we described the situation, when the spin is initially oriented at $t=0$, and described the following spin dynamics in the external magnetic field. Experimentally, the spin initialization and measurement are repeated many times to increase the signal to noise ratio. So the spin is excited by a train of pump pulses. If the repetition period of the pulses $T_R$ exceeds the spin relaxation time $\tau_s$, it is enough to study theoretically the spin dynamics after a single pulse only, which was described in the previous sections. In the alternative approach, when the spin is pumped continuously, the average in Eq.~\eqref{eq:average_spin} describes the steady state spin polarization.

In this section we consider the case of the pulsed excitation, when the repetition period is comparable with the spin lifetime (for the isotropic hyperfine interaction and frozen nuclear spins).

\subsection{Spin dynamics}
\label{sec:pulses_general}

\begin{figure}
  \centering
  \includegraphics[width=0.75\linewidth]{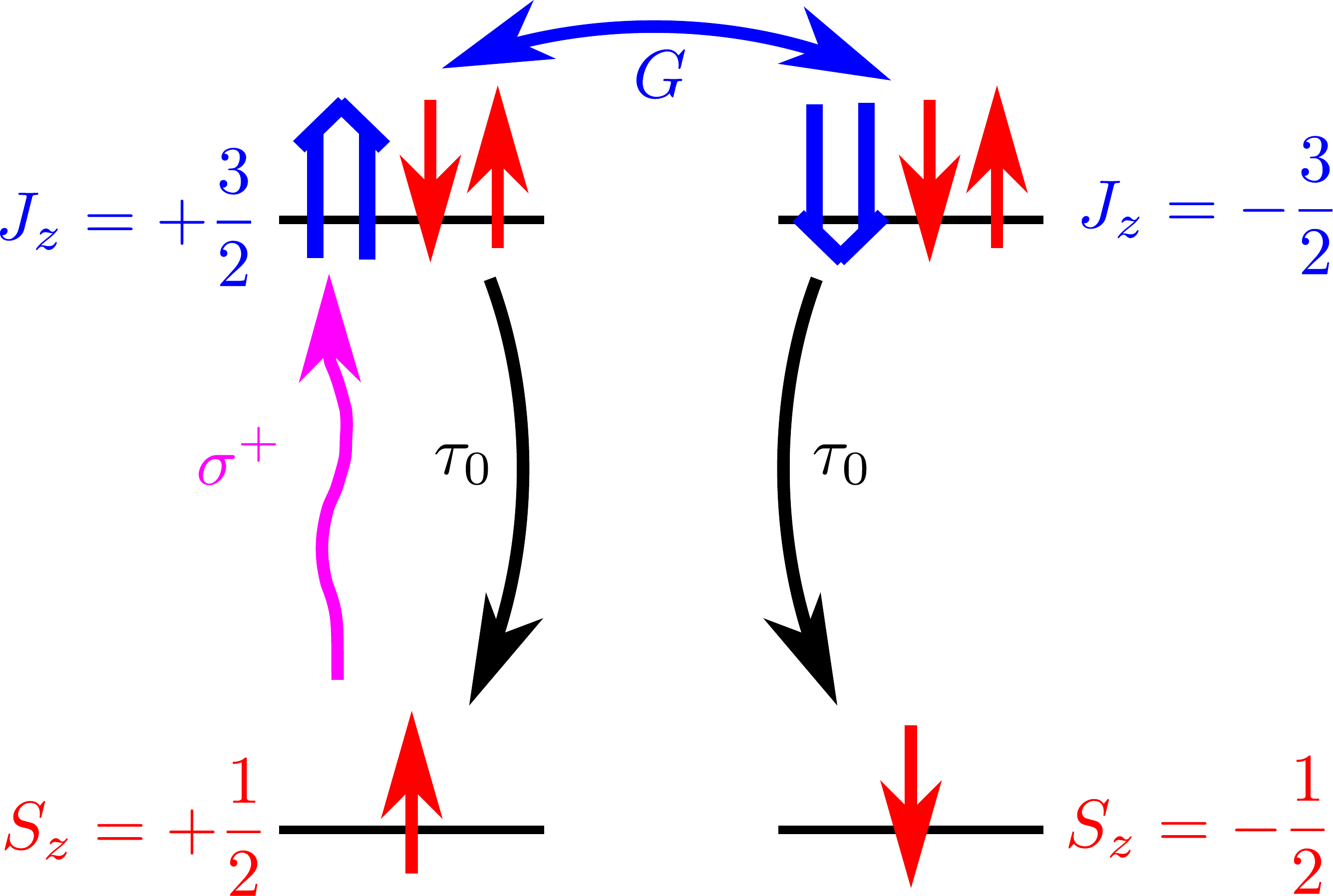}
  \caption{Energy levels in the QD and transitions between them. The red and blue arrows denote the electron and heavy hole spins, respectively. The excitation with $\sigma^+$ polarized light is shown by the magenta arrow, black arrows denote the trion recombination, and the blue arrow at the top shows the trion spin relaxation, which leads to the spin generation in the ground state.}
  \label{fig:trions}
\end{figure}

Let us consider the resonant electron spin pumping, when the photon energy of the pump pulse coincides with the optical resonance of the singlet heavy-hole trion. We remind that the negatively charged trion is composed of  two electrons with opposite spins and a hole. We assume that the trion lifetime $\tau_0$ (typically, of about $1$~ns~\cite{GreilichPRB73}) is much shorter than the repetition period of the pump pulses $T_R$. In this case, there are no trions at the moments of the arrivals of the pump pulses. The pump pulses create trions accordingly to the optical selection rules~\cite{ivchenko05a}. Namely, $\sigma^\pm$ photons are absorbed only for the electron spin $S_z=\pm1/2$, respectively, as illustrated in Fig.~\ref{fig:trions}. The electron spin after the trion recombination (not immediately after the pump pulse), $\bm S^+$, is related with the spin before the pulse, $\bm S^-$, as~\cite{yugova09,PhysRevResearch.1.033189}
\begin{subequations}
  \label{eq:pp}
  \begin{equation}
    \label{eq:Delta_Sz}
    S_z^+=S_z^-+\frac{1-Q^2}{2}\left(\frac{\mathcal P}{2}-S_z^-\right)G,
  \end{equation}
  \begin{equation}
    \label{eq:Delta_Sxy}
    S_x^+=QS_x^-,
    \quad
    S_y^+=QS_y^-.
  \end{equation}
\end{subequations}
Here $Q\in[-1;1]$ is the amplitude of the probability \textit{not} to excite a trion, $\mathcal P=\pm1$ stands for the helicity of the pump pulse, and $G\in[0;1]$ is the spin generation efficiency, which accounts for the trion spin dynamics~\cite{PhysRevB.98.125306}. The $\pi$-pulses are described by $Q=0$, in this case the probability of the trion excitation is $1$.

The second term in the right hand side of Eq.~\eqref{eq:Delta_Sz} describes the change of the spin polarization due to the trion excitation with the probability $1-Q^2$. The brackets describe the saturation of the spin polarization at the value $\mathcal P/2$. The parameter $G/2$ equals to the trion spin flip probability. If the trion spin relaxation is absent ($G=0$), then the trion excitation and recombination does not change spin in the ground state, as can be seen in Fig.~\ref{fig:trions}. However, if the trion spin relaxes before the trion recombination, $G=1$, the pump pulse creates spin polarization. The $\pi$ pulse in this case depolarizes electrons in one spin state, and leaves the other spin state untouched. This results in the spin $S_z^+=\mathcal P/4$ after the pulse which is the upper limit for the spin polarization after a single pulse.

From Eq.~\eqref{eq:Delta_Sxy} one can also see that the trion excitation destroys the in plane spin components~\cite{zhukov10}. If the pump pulses were slightly detuned from the trion resonance the electron spins would be additionally rotated around the $z$ axis. The strongly nonresonant spin pumping can be phenomenologically described by Eqs.~\eqref{eq:pp} as well.

Between the pump pulses we assume that the electron spin precesses in the frozen nuclear field and external magnetic field and relaxes with the time $\tau_s\gg1/\delta$:
\begin{equation}
  \label{eq:Bloch2}
  \frac{\d\bm S}{\d t}=\bm\Omega_{\rm tot}\times\bm S-\frac{\bm S}{\tau_s},
\end{equation}
as described in Sec.~\ref{sec:basic}. We remind that $\bm\Omega_{\rm tot}=\bm\Omega_N+\bm\Omega_L$ is the total electron spin precession frequency. Under the long pulsed excitation, the steady state spin dynamics is established. We assume that the spin polarization is probed shortly before the pump pulses (this is equivalent to the delay a bit shorter than $T_R$), so we aim at the calculation of $\braket{S_z^-}$. To find it we solve the Bloch equation~\eqref{eq:Bloch2} with the initial condition $\bm S^+$. In the steady state, after the time $T_R$ the solution has to coincide with $\bm S^-$~\cite{efros03,A.Greilich07212006,yugova09}. From this relation along with Eq.~\eqref{eq:pp} we find $S_z^-$ and then average it over the nuclear field distribution function~\eqref{eq:F}. Below we describe the dependence of $\braket{S_z^-}$ on the magnetic field.

\subsection{Hanle and PRC for pulsed excitation}\label{subsec:pulsed_excitation}

The spin dynamics under pulsed excitations was extensively studied in transverse magnetic field, see Ref.~\onlinecite{yugova12} for a summary. The most spectacular effects in this configuration are resonant spin amplification~\cite{Kikkawa98} and spin mode locking~\cite{A.Greilich07212006}. In this paper we will not reproduce these results, and will pay the main attention to the Faraday geometry.%, when the longitudinal magnetic field is applied to the system. In this geometry the pulsed resonant trion excitation can lead to new effects, which we describe below.

The spin dynamics under pulsed excitation of trions in the longitudinal magnetic field is largely unexplored~\cite{PhysRevB.98.125306}. Here we will consider only the most experimentally relevant limit of long spin relaxation time and small average spin polarization:
\begin{equation}
  \label{eq:assum}
  \tau_s\delta\gg1,
  \qquad
  G\tau_s/T_R\ll 1.
\end{equation}
In this limit the shapes of the Hanle curve and PRC strongly depend on the repetition period of the pump pulses, $T_R$, and on the strength of the pulses, $Q$.

To establish the relation with the initial spin polarization $S_0$ used in the previous sections, let us consider the limit of large longitudinal magnetic field, $\Omega_L\gg\delta,1/T_R$. In this limit the hyperfine interaction plays no role, so Eq.~\eqref{eq:Bloch2} simply yields
\begin{equation}
  \bm S^-=\bm S^+\exp(-T_R/\tau_s).
\end{equation}
Combining this with Eqs.~\eqref{eq:pp} we find
\begin{subequations}
  \begin{equation}
    \label{eq:S0}
    S_0\equiv S_z^-=\frac{(1-Q^2)\mathcal P G}{4[\exp(T_R/\tau_s)-1]}\ll 1,
  \end{equation}
  \begin{equation}
    S_x^-=S_y^-=0.
  \end{equation}
\end{subequations}
This naturally shows that the longer the repetition period the smaller the polarization. The condition~\eqref{eq:assum} indeed results in the small maximum spin polarization $S_0$. The case of $S_0\sim1$ was studied in Ref.~\onlinecite{PhysRevResearch.1.033189}.

\begin{figure}
  \centering
  \includegraphics[width=\linewidth]{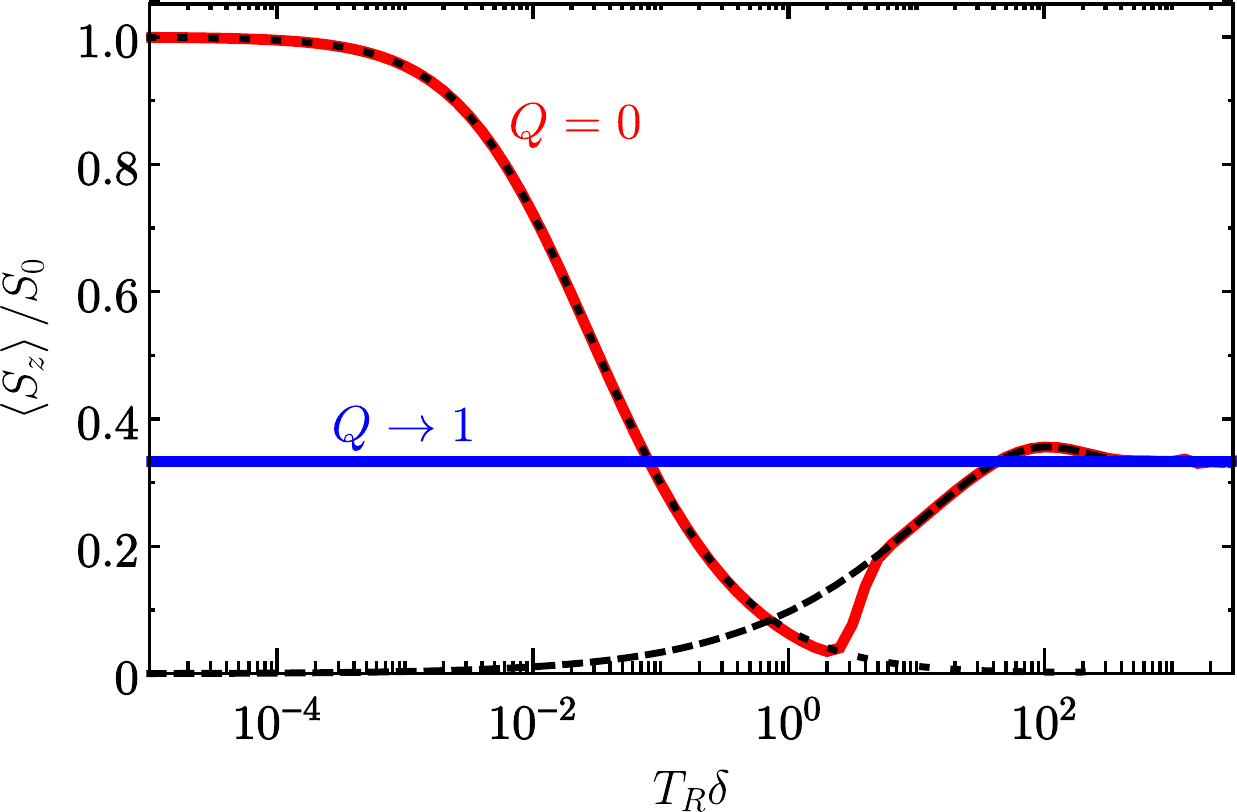}
  \caption{The dependence of the spin polarization on the repetition period of the pump pulses for weak pulses, $Q\to 1$ (blue curve) [Eq.~\eqref{eq:Q1}], and strong pulses, $Q=0$ (red curve). The black dotted and dashed curves show the approximate Eqs.~\eqref{eq:small_TR} and~\eqref{eq:large_TR}, respectively. The calculations are performed with $\tau_s\delta=100$.}
  \label{fig:TR_dep}
\end{figure}

Now let us consider the spin polarization in zero magnetic field, $\Omega_L=0$. It is shown in Fig.~\ref{fig:TR_dep} as a function of $T_R$. In the limit of weak pulses, $Q\to 1$ (blue curve), the spin polarization is constant:
\begin{equation}
  \label{eq:Q1}
  \braket{S_z}=S_0/3.
\end{equation}
We recall that $\braket{S_z}$ denotes in this section the average value of $S_z^-$ in the steady state. In this limit the model of Sec.~\ref{sec:basic} is valid.
% This is related with the fact, that the spin polarization is probed just before the pump pulse arrival, which is equivalent to the delay $T_R$. Therefore, the probed spin polarization is simply given by
% \begin{equation}
%   \label{eq:Q1}
%   \frac{\braket{S_z}}{S_0}=\frac{1}{3}\frac{T_R/\tau_s}{\e^{T_R/\tau_s}-1}.
% \end{equation}

The spin polarization for strong pulses, $Q=0$, is shown by the red curve in Fig.~\ref{fig:TR_dep}. With increase of the repetition period it decays from $S_0$ almost to zero at $T_R\sim1/\delta$ and then increases to $S_0/3$. For the long repetition period, $T_R\gg1/\delta$, the situation is similar to the previous case, because the spin polarization almost completely decays between the pump pulses.

For the short repetition period, $T_R\ll1/\delta$, each pump pulse projects the spin polarization on the $z$ axis, as follows from Eqs.~\eqref{eq:pp}. Between the pump pulses, the spin precesses in the nuclear field, $\bm \Omega_N$, but the typical rotation angle is small. This results in the effective spin relaxation between the pump pulses, which is described by the relation
\begin{multline}
  \label{eq:Sz_din_short}
  S_z^-=S_z^+\cos(\Omega_{N,\perp}T_R)\exp(-T_R/\tau_s)\\\approx S_z^+\left(1-\frac{(\Omega_{N,\perp}T_R)^2}{2}-\frac{T_R}{\tau_s}\right),
\end{multline}
where $\Omega_{N,\perp}$ is the component of the nuclear field in the $(xy)$ plane. One can see that the effective spin relaxation rate in this limit is
\begin{equation}
  \label{eq:tau_s_eff2}
  \frac{1}{\tau_{\rm eff}}=\frac{1}{\tau_s}+\frac{\Omega_{N,\perp}^2T_R}{2},
\end{equation}
and it increases with an increase of the repetition period. Note the difference between the effective spin relaxation time in this case and in the case of fast nuclear spin dynamics, Eq.~\eqref{eq:tau_s_eff}. Averaging over the nuclear field distribution function yields in this limit the analytical expression
\begin{equation}
  \label{eq:small_TR}
  \frac{\braket{S_z}}{S_0}=-\nu\exp(\nu)\Ei(-\nu),
\end{equation}
where $\nu=2/(T_R\tau_s\delta^2)$ and $\Ei(x)=-\int_{-x}^\infty\exp(-t)/t\d t$ is the exponential integral function. The spin polarization decreases with increase of $T_R$, as shown by the black dotted curve in Fig.~\ref{fig:TR_dep}.

Curiously, this dependence can be viewed as the manifestation of the quantum Zeno effect~\cite{khalfin1958contribution,doi:10.1063/1.523304}. Indeed, the random nuclear field leads to the electron spin precession and dephasing. However, each pump pulse acts as a measurement and projects the spin on the $z$ direction. Actually, absorption of the pump pulses can be in principle measured experimentally and this would give the value of $S_z^-$. The fast measurements (short repetition period) freeze the spin dynamics and increase the effective spin relaxation time as described by Eq.~\eqref{eq:tau_s_eff2}. Moreover, if the measurement rate is comparable with the spin precession frequency, $T_R\delta\sim1$, the spin relaxation rate increases and becomes faster than in the absence of the measurements. The spin polarization in this case drops to zero, and this is known as the quantum anti-Zeno effect~\cite{LANE1983359,Facchi_2008}.

In the opposite limit of long repetition period, ${T_R\gg1/\delta}$, the spin rotates by the large angle $\sim T_R\delta$ between the two pump pulses, so one can average the spin polarization $S_z^-$ over it and obtain
\begin{multline}
  \label{eq:large_TR}
  \frac{\braket{S_z}}{S_0}=1-\exp(T_R/\tau_s)\\
  +\sqrt{\frac{\exp(T_R/\tau_s)-1}{2}}\arctg\left(\sqrt{\dfrac{2}{\exp(T_R/\tau_s)-1}}\right).
\end{multline}
% \begin{equation}
%   \label{eq:large_TR}
%   \frac{S_z}{S_0}=1-\e^{T_R/\tau_s}+\sqrt{\frac{\e^{T_R/\tau_s}-1}{2}}\arctg\left(\sqrt{\dfrac{2}{\e^{T_R/\tau_s}-1}}\right).
% %   \frac{S_z}{S_0}=\left[
% % -1+\frac{\e^{T_R/\tau_s}\arctg\left(\sqrt{\dfrac{2}{\e^{T_R/\tau_s}-1}}\right)}{\sqrt{2\left(\e^{T_R/\tau_s}-1\right)}}
% % \right](\e^{T_R/\tau_s}-1).
% \end{equation}
This expression is shown by the black dashed curve in Fig.~\ref{fig:TR_dep}. In particular, with increase of $T_R$ it increases at $T_R\ll\tau_s$ as $(\pi/2)\sqrt{T_R/(2\tau_s)}$ and saturates at $T_R\gg\tau_s$. Qualitatively, this expression describes the smooth transition from Eq.~\eqref{eq:small_TR} to Eq.~\eqref{eq:Q1}.

\begin{figure}
  \centering
  \includegraphics[width=\linewidth]{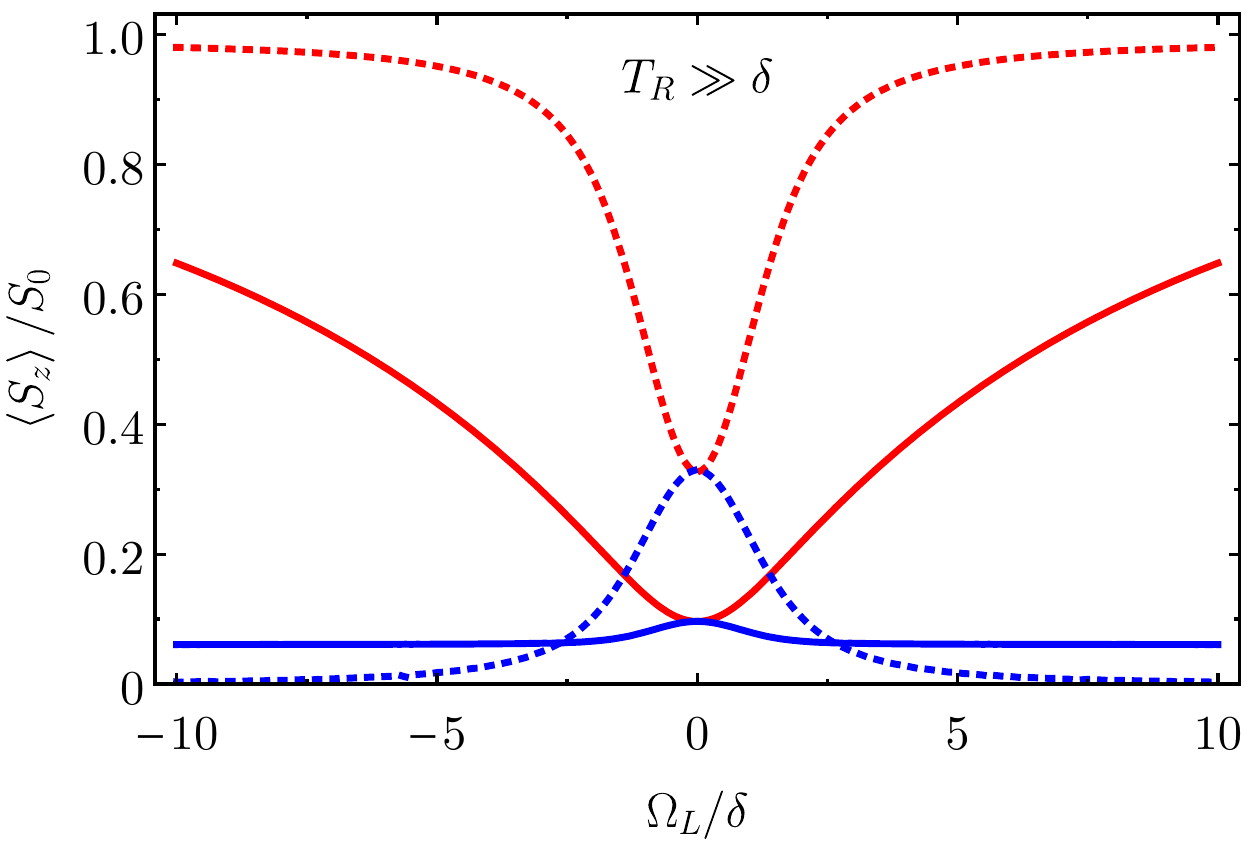}
  \caption{The PRC (red) and Hanle curve (blue) for the parameters $\tau_s\delta=1000$, $T_R\delta=10$ for weak, $Q\to1$, (dotted curves) and $\pi$ pump pulses, $Q=0$, (solid curves).}
  \label{fig:cont_pulses}
\end{figure}

The PRC and Hanle curve for the long repetition period, $T_R\gg1/\delta$ are shown in Fig.~\ref{fig:cont_pulses} for weak and strong pulses. In the limit of weak pulses, $Q=1$, the dotted curves are described by Eqs.~\eqref{eq:isotropic}. In the case of the strong pulses, $Q=0$, (solid curves) the spin polarization gets suppressed (relative to $S_0$), as expected from Eqs.~\eqref{eq:Delta_Sxy}. Moreover, the PRC broadens considerably~\cite{PhysRevResearch.1.033189}. This is related with the fact that the strong pump pulses ``erase'' the transverse spin components, see Eqs.~\eqref{eq:Delta_Sxy}, so the total spin polarization decreases even in quite strong magnetic field. In addition, the offset appears in the Hanle curve (it does not decay to zero), which is a manifestation of the spin mode locking effect~\cite{A.Greilich07212006,yugova12}. %\commentDima{Describe PRC analytically?}

\begin{figure}
  \centering
  \includegraphics[width=\linewidth]{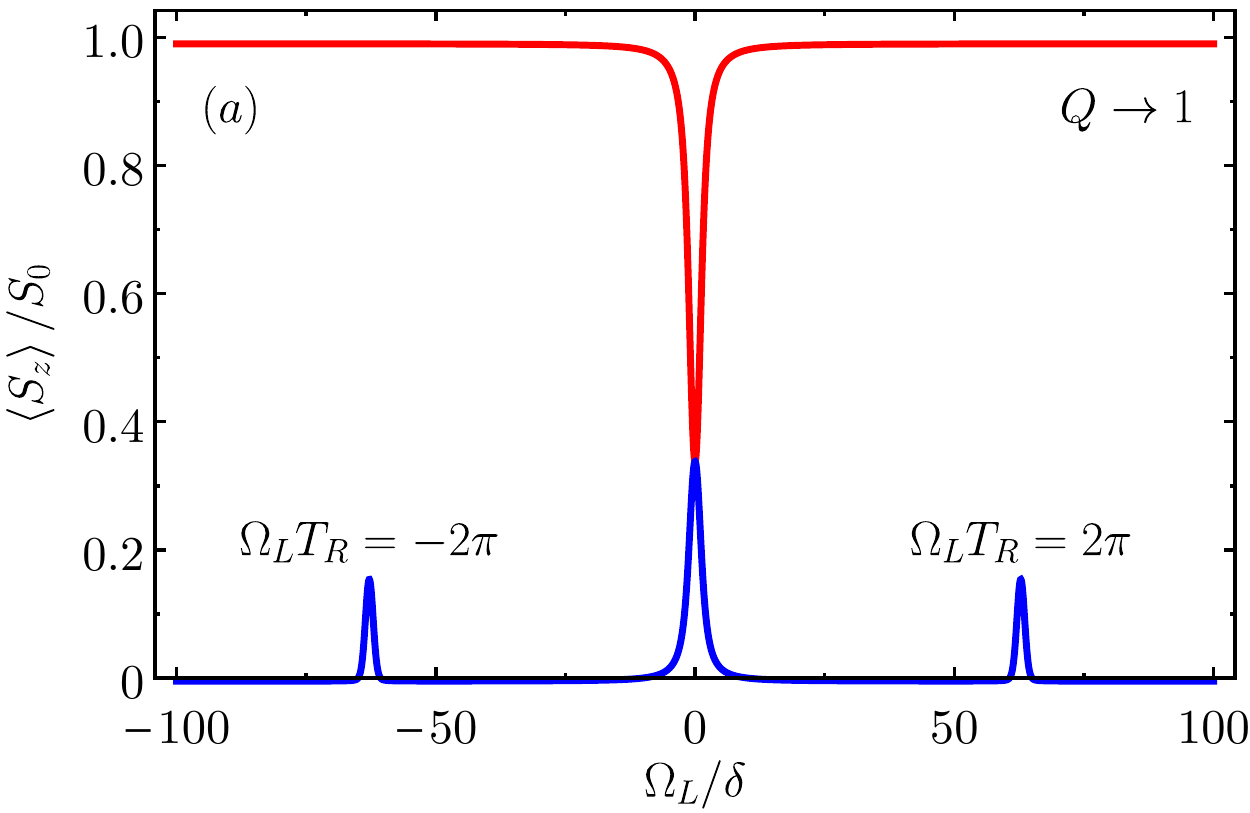}\\[2mm]
  \includegraphics[width=\linewidth]{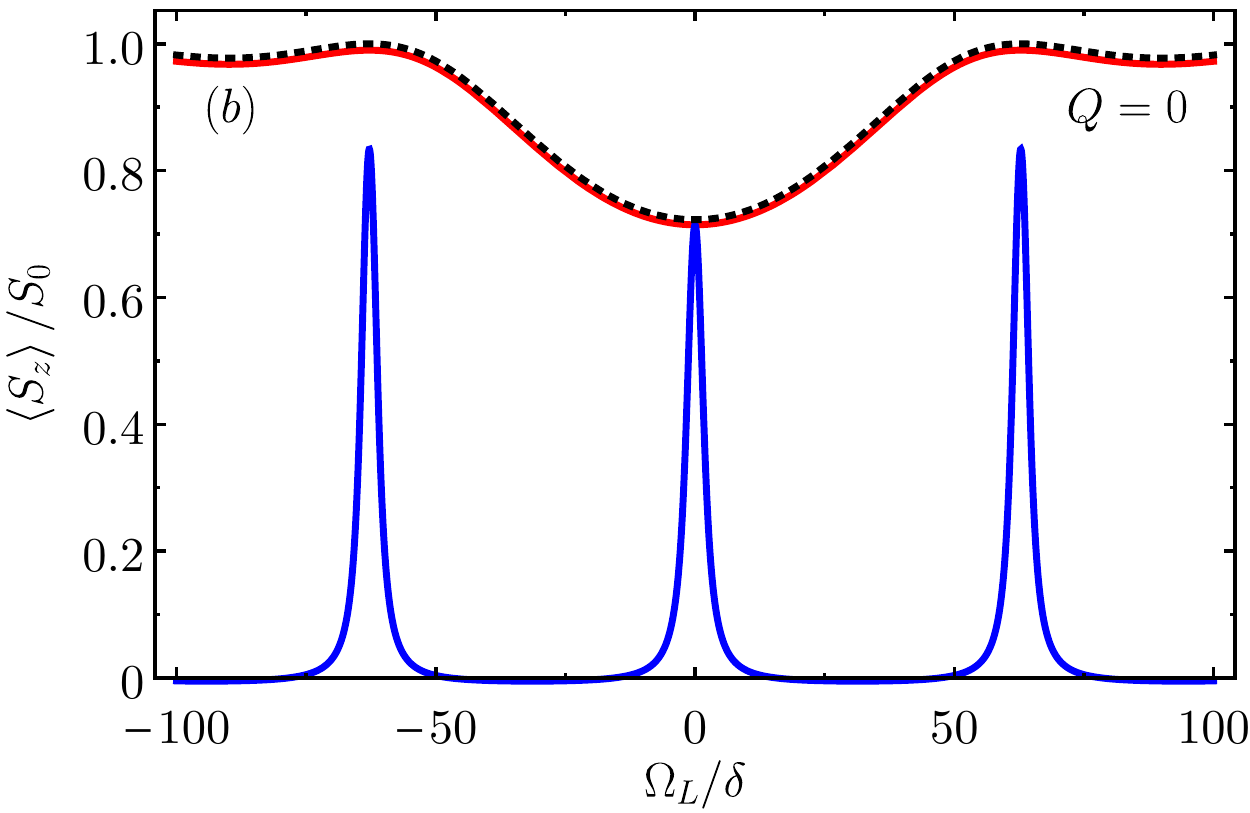}
  \caption{PRC (red) and Hanle curve (blue) for the short repetition period, $T_R\delta=0.1$, for (a) weak, $Q\to1$, and (b)~strong, $Q=0$, pump pulses calculated for $\tau_s\delta=10$. The black dotted curve in (b) shows Eq.~\eqref{eq:small_TR} with account for Eq.~\eqref{eq:tau_s_P}.}
  \label{fig:short_TR}
\end{figure}

The PRC and Hanle curve for the short repetition period, $T_R \ll 1/\delta$, are shown in Fig.~\ref{fig:short_TR}. For the weak pulses, $Q\to1$, the shape of the Hanle curve at small magnetic fields is described by Eq.~\eqref{eq:isotropic_H}, while at the magnetic fields corresponding to $\Omega_L T_R=\pm 2\pi$ one can see resonant spin amplification (RSA)~\cite{Awschalom_Spintronics,Kikkawa98,yugova12}. For the strong pump pulses, $Q=0$, the dependence on the transverse magnetic field is similar except for the larger spin polarization, as described by Eq.~\eqref{eq:small_TR}. In the Faraday geometry, the PRC for weak pulses is described by Eq.~\eqref{eq:isotropic_P}.

The most interesting is the PRC for strong pump pulses, red solid curve in Fig.~\ref{fig:short_TR}(b). One can see that it is much broader than $\delta$ and has the width of the order of $1/T_R$. The reason is that, one has to apply the strong magnetic field $\Omega_L\sim1/T_R$ in order to suppress the spin relaxation described by Eq.~\eqref{eq:small_TR}. In this case the electron spin rotates around the axis slightly tilted from the $z$ axis by the transverse components of the nuclear field. As a result instead of Eq.~\eqref{eq:Sz_din_short} we obtain
\begin{equation}
  \frac{S_z^-}{S_z^+}=1-\frac{\Omega_{N,\perp}^2}{\Omega_L^2}\left[1-\cos(\Omega_LT_R)\right].
\end{equation}
This again results in the renormalized spin relaxation rate, and Eq.~\eqref{eq:small_TR} is still valid provided that
\begin{equation}
  \label{eq:tau_s_P}
  \nu(\Omega_L)=\frac{\Omega_L^2T_R}{\tau_s\delta^2\left[1-\cos(\Omega_LT_R)\right]}.
\end{equation}
This expression is shown in Fig.~\ref{fig:short_TR}(b) by the black dotted curve. Note the weak oscillations with the same period as RSA, which we describe below.

\begin{figure}
  \centering
  \includegraphics[width=\linewidth]{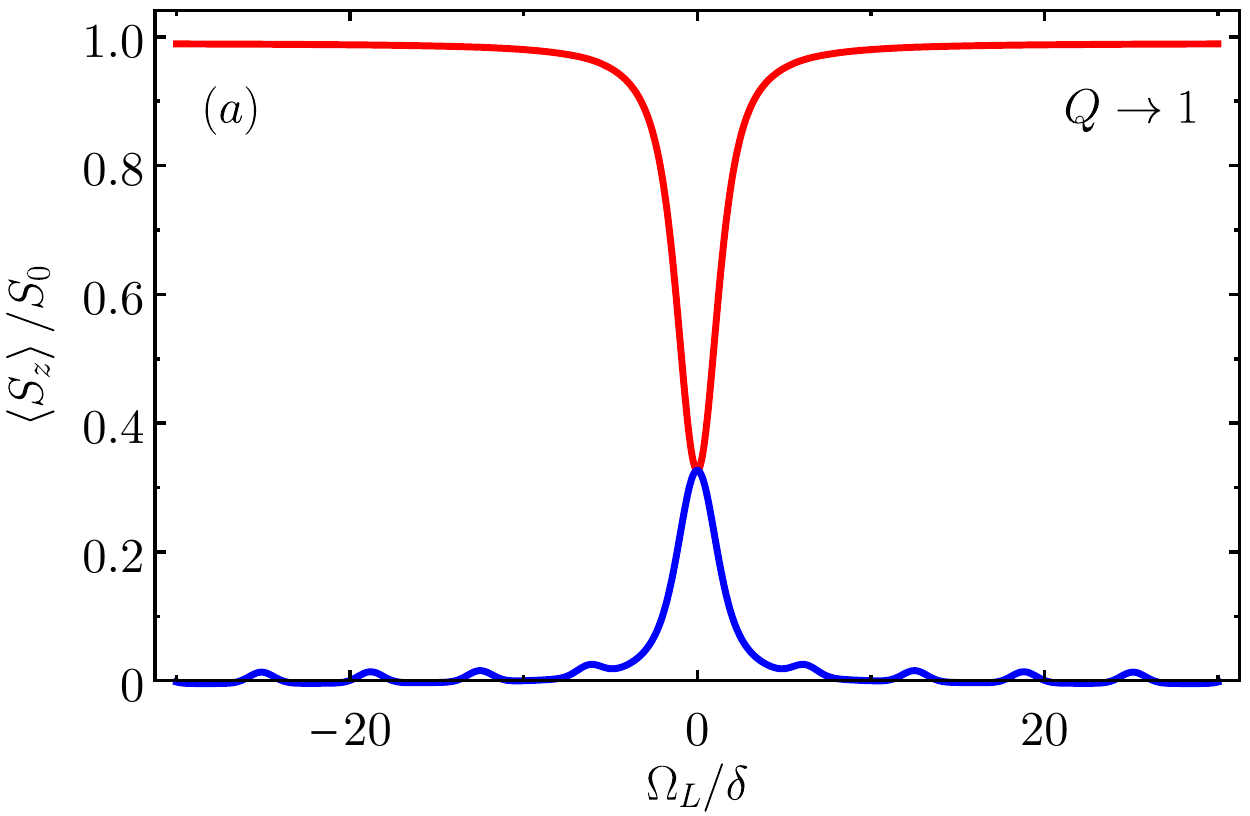}\\[2mm]
  \includegraphics[width=\linewidth]{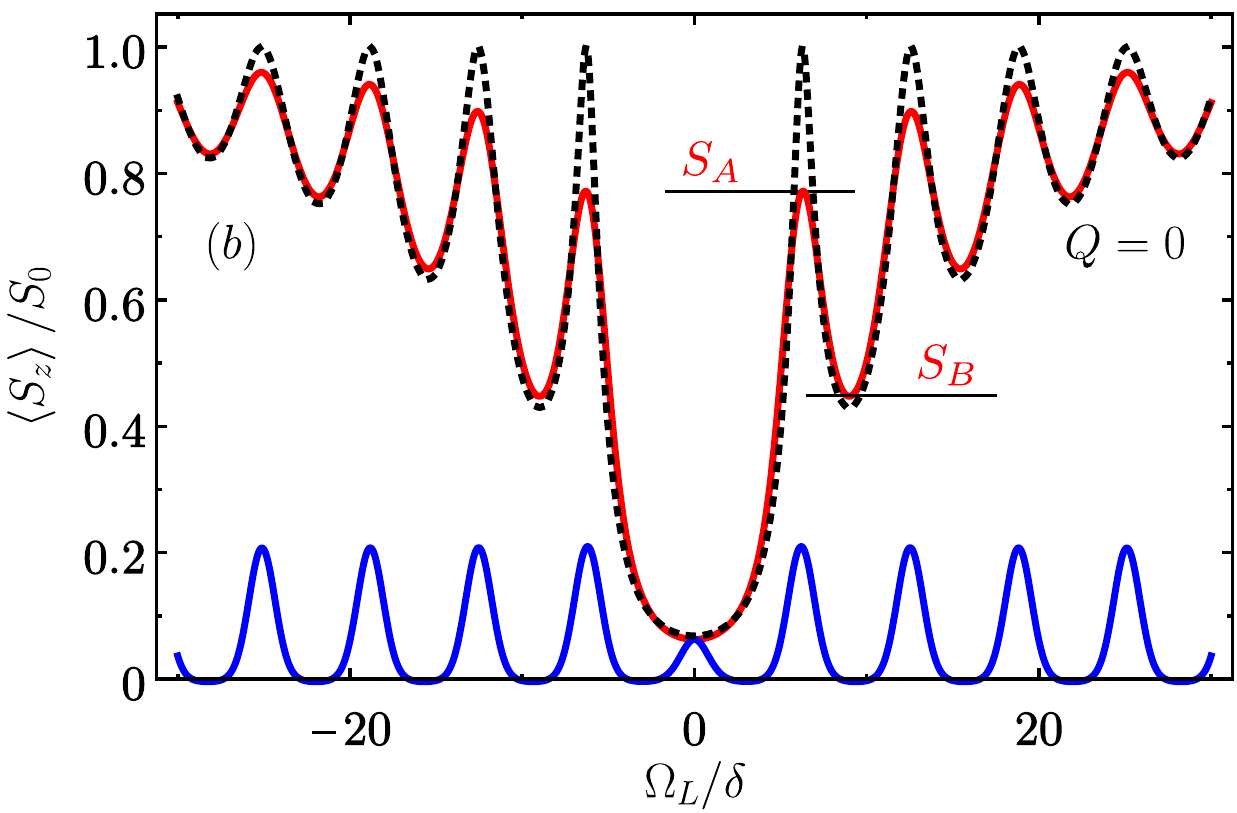}
  \caption{The same as in Fig.~\ref{fig:short_TR}, but for $T_R\delta=1$ (and $\tau_s\delta=100$).}
  \label{fig:Zeno_pulses}
\end{figure}

\begin{figure}
  \centering
  \includegraphics[width=\linewidth]{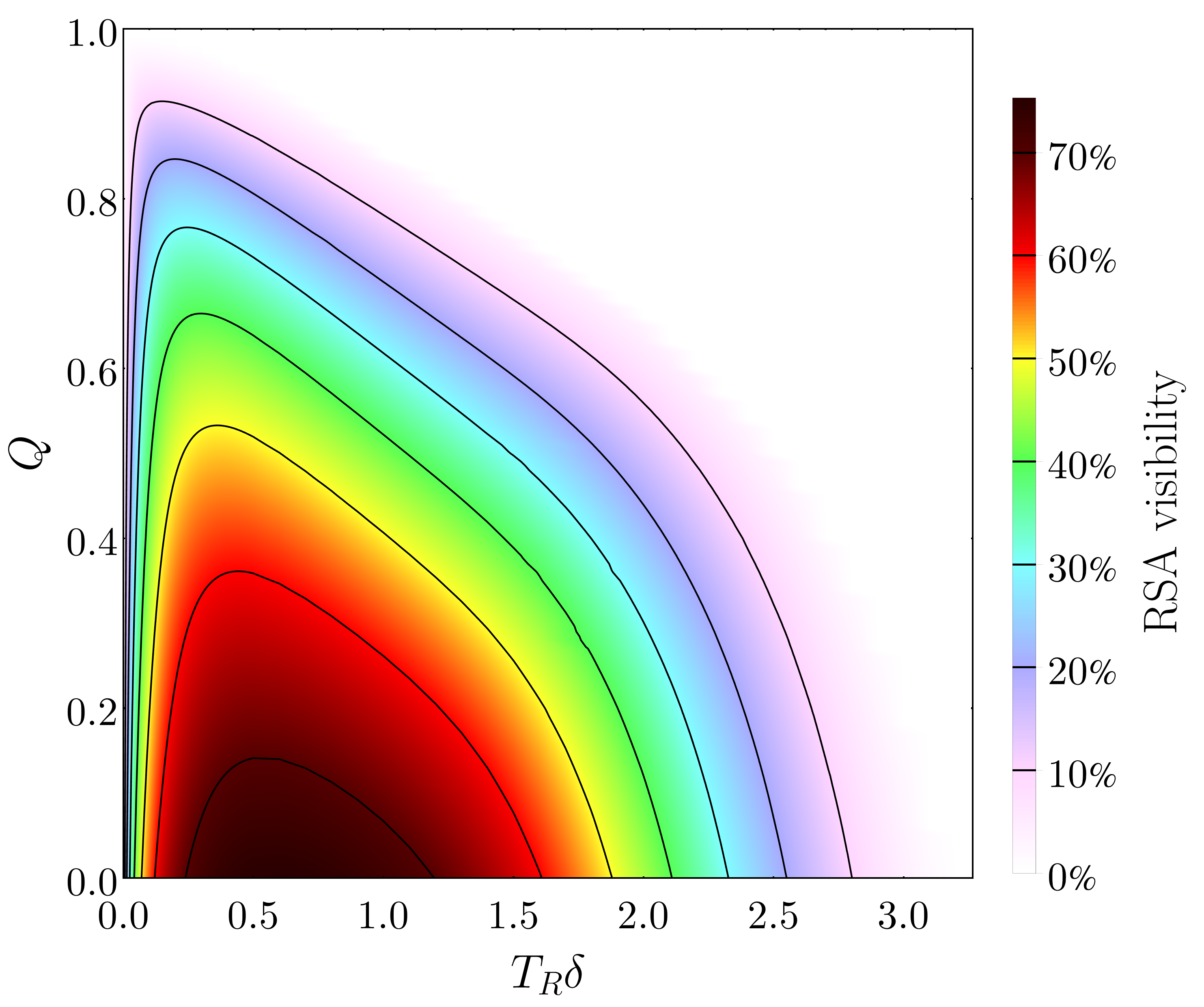}
  \caption{The visibility of RSA in the Faraday geometry, Eq.~\eqref{eq:V}, calculated for $\tau_s\delta=100$.}
  \label{fig:visibility}
\end{figure}

Finally, the PRC and Hanle curve for the intermediate repetition period, $T_R\delta=1$, are shown in Fig.~\ref{fig:Zeno_pulses}. Generally, all the dependencies here are similar to Fig.~\ref{fig:short_TR}, but the oscillations in the PRC for strong pump pulses, $Q=0$, are very pronounced. They can be called RSA in the Faraday geometry~\cite{PhysRevResearch.1.033189}. To analyze them in more detail we introduce the visibility $V$, defined as the ratio of the difference between the first maximum, $S_A$, and the first minimum, $S_B$ to the first maximum:
\begin{equation}
  \label{eq:V}
  V=\frac{S_A-S_B}{S_A},
\end{equation}
see Fig.~\ref{fig:Zeno_pulses}(b). The visibility is shown in Fig.~\ref{fig:visibility} as a function of the repetition period and the pump pulse power. One can see that it is considerable only for the strong pump pulses, and only in the intermediate range of the repetition periods, $T_R\delta\sim1$. For the long and short repetition periods RSA in the Faraday geometry vanishes, as shown in Figs.~\ref{fig:cont_pulses} and~\ref{fig:short_TR}. Note that with decrease of the spin relaxation time $\tau_s$ the visibility also decreases, and becomes small already for $\tau_s\delta\lesssim 10$. However, for long spin relaxation times, the visibility can exceed $70\%$ as shown in Fig.~\ref{fig:visibility}, so the oscillations are very pronounced.

Experimentally, RSA in Faraday geometry was not observed so far. In typical experiments $T_R=13.2$~ns and $\delta=0.44$~ns$^{-1}$~\cite{Glasenapp2016,PhysRevB.98.121304}, so $T_R\delta\approx 6$. For holes the hyperfine interaction is weaker, but it is also strongly anisotropic, see Sec.~\ref{sec:Anisotrop}, and RSA in this case is strongly suppressed~\cite{PhysRevResearch.1.033189}. We believe that decrease of the pulse repetition period for $n$-doped QDs and increase of the power of pump pulses will allow one to observe this effect.

\subsection{Effect of trion spin dynamics}\label{subsec:M-form}

In the previous subsection we implicitly assumed that the spin generation rate $G$ does not depend on magnetic field (the ratio $\braket{S_z}/S_0$ did not depend on $G$). In reality, it is determined by the trion spin dynamics~\cite{PhysRevB.98.125306}, as discussed in Sec.~\ref{sec:pulses_general}. The trion spin $\bm J$ obeys the equation
\begin{equation}
  \label{eq:J_Bloch}
  \frac{\d\bm J}{\d t}=\left(\bm\Omega_N^{\rm T}+\bm\Omega_L^{\rm T}\right)\times\bm J-\frac{\bm J}{\tau_s^{\rm T}}-\frac{\bm J}{\tau_0},
\end{equation}
which is similar to Eq.~\eqref{eq:Bloch}, but with the parameters relevant for the trion spin dynamics, which we denote by the superscript ``$\rm T$''. Thus, $\tau_s^T$ refers to the trion spin relaxation time. We recall that $\tau_0$ is the trion lifetime. To be specific, we focus here on the negatively charged trion, but the same formalism can be also applied to the positively charged trions (see Sec.~\ref{sec:M-shape}). The trion consists of two electrons in the singlet spin state and a heavy hole with unpaired pseudospin $\bm J$, see Fig.~\ref{fig:trions}. Therefore, the spin dynamics of the trion are related with the hyperfine interaction and Zeeman splitting of the heavy hole. It should be noted that the $g$ factor and the hyperfine interaction for holes are usually strongly anisotropic, so that $g_{h,\parallel}\gg g_{h,\perp}$ and $\delta_{x,y} \ll \delta_z$.

\begin{figure}
  \includegraphics[width=\linewidth]{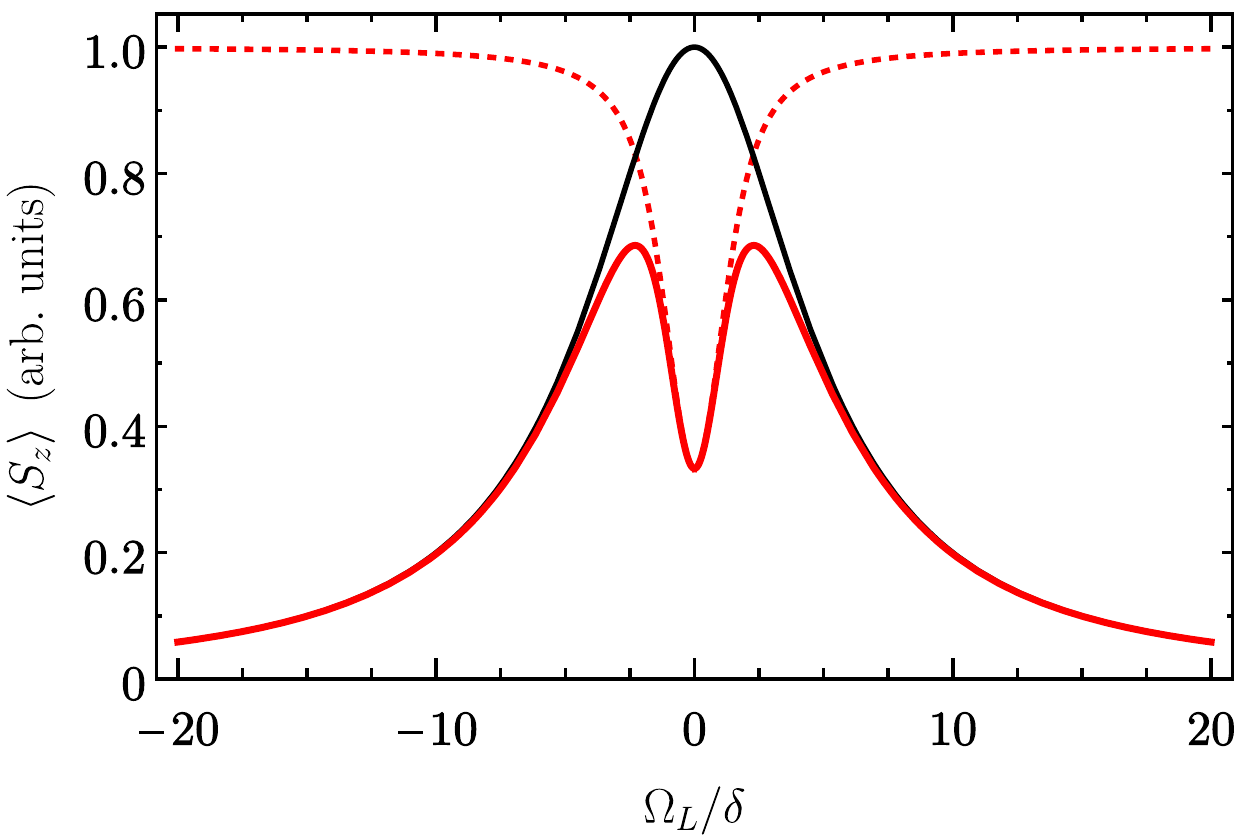}
  \caption{PRC (solid red curve) calculated after Eq.~\eqref{eq:S0P} for the parameters $\delta_{\rm T}=\delta$, $\Omega_L^{\rm T}=\Omega_L$, $\tau_0\delta=0.2$ in the limit $\delta_{\rm T}^2\tau_s^{\rm T}\tau_0\gg 1$. The red dashed and black solid curves show $P(\Omega_L)$ [Eq.~\eqref{eq:isotropic_P}] and $G(\Omega_L^{\rm T})/G(0)$ [Eq.~\eqref{eq:G}], respectively.}
  \label{fig:M-shape}
\end{figure}

The trion spin flip probability during its lifetime is defined similarly to Eq.~\eqref{eq:average_spin}~\cite{PhysRevB.98.125306}:
\begin{equation}
  G(\Omega_L^{\rm T})=1-\frac{1}{J_0\tau_0}\int\limits_0^\infty J_z(t)\d t,
\end{equation}
where $\bm J(0)=J_0\bm e_z$. Typically, the trion lifetime $\tau_0$ is shorter than the trion spin relaxation $\tau_s^{\rm T}$ and the typical trion spin precession period $1/\delta_{\rm T}$, where the parameter $\delta_{\rm T}$ is analogous to $\delta$, but for the trion state (for simplicity, we neglect the anisotropy). For the resident electrons under study, $\delta_{\rm T}$ describes the hyperfine interaction of the heavy hole in the singlet trion state (while for resident holes the situation is reversed). In this case, the solution of Eq.~\eqref{eq:J_Bloch} averaged over $\bm\Omega_N^{\rm T}$ in the longitudinal magnetic field yields
\begin{equation}
  \label{eq:G}
  G(\Omega_L^{\rm T})=\frac{\tau_0}{\tau_s^{\rm T}}+\frac{(\delta_{\rm T}\tau_0)^2}{1+(\Omega_L^{\rm T}\tau_0)^2},
\end{equation}
c.f. Eq.~\eqref{eq:tau_s_eff}. Here the first term describes the trion spin flip probability associated with $\tau_s^{\rm T}$ and the second term describes the nuclei related trion spin relaxation. In the limit $\tau_s^{\rm T}\ll1/(\delta_{\rm T}^2\tau_0)$, the trion spin relaxation is not dominated by the hyperfine interaction, so the spin generation rate does not depend on the magnetic field, as was assumed in the previous sections. In the opposite limit, the generation rate gets suppressed by the magnetic field, but this suppression requires relatively strong magnetic field $\Omega_L^{\rm T}\gtrsim1/\tau_0$ related to the short trion lifetime. This dependence is shown in Fig.~\ref{fig:M-shape} by the black solid curve. For the weak pump pulses $Q\to 1$, the total spin polarization as a function of the magnetic field is described by
\begin{equation}
  \label{eq:S0P}
  \braket{S_z}=S_0P(\Omega_L),
\end{equation}
where $S_0$ is given by Eq.~\eqref{eq:S0} and $P(\Omega_L)$ is given by Eq.~\eqref{eq:isotropic_P}.

Application of the longitudinal magnetic field suppresses trion spin relaxation and, as a result, the resident charge carrier spin pumping, see Eq.~\eqref{eq:Delta_Sz}. The corresponding PRC has M-like shape, which is shown in Fig.~\ref{fig:M-shape} by the red solid line. Here the spin polarization $\braket{S_z}$ is plotted in contrast with the previous figures, because $S_0$ in this case depends on $\Omega_L$. For this calculation the first term in Eq.~\eqref{eq:G} is neglected. If it is comparable to the second term, then the spin polarization does not decay to zero, but saturates in the large magnetic fields.

In the rest of the paper we present experimental results, which illustrate the theoretical models.

\section{Experimental details}

\subsection{Samples}\label{subsec:samples}

In this paper we use a set of four samples to illustrate the various mechanisms controlling PRC and Hanle curve: (i) Sample-1 is an example for a basic type of Hanle and PRC, as dicussed in Sec.~\ref{sec:basic}. (ii) Samples 2 and 3 demonstrate the effect of the nuclear spin correlation time, see Sec.~\ref{sec:Nuclear}. (iii) Sample-3 presents an additive contribution for two types of carriers (electrons and holes). (iv) Sample-4 shows M-shape of the PRC, as discussed in Sec.~\ref{subsec:M-form}, and additionally demonstrates the effect of the anisotropic hyperfine interaction, Sec.~\ref{sec:Anisotrop}.

Sample-1 is the antireflection coated \textit{n}-type GaAs epilayer sample of 350~$\mu$m thickness and a donor of $n_D=1.4\times10^{16}$~cm$^{-3}$. For details see Ref.~\cite{Crooker2009}.

All the other samples are grown by the molecular-beam epitaxy (MBE) on $(100)$-oriented GaAs substrates.

Sample-2 (\#2018) is an \textit{n}-type ZnSe/Zn$_{0.85}$Mg$_{0.15}\text{Se}$ single quantum well (QW) structure. The sample starts with a 3.4-nm-thick ZnSe layer, which is used to reduce the strain induced by the II-VI / III-V heterointerface. It is followed by a 24-nm-thick Zn$_{0.85}$Mg$_{0.15}\text{Se}$ barrier layer, which prevents carrier diffusion into the substrate. The 20-nm-thick undoped ZnSe QW is grown on top of the barrier layer, followed by the 30-nm-thick Zn$_{0.85}$Mg$_{0.15}\text{Se}$ upper barrier. The QW is nominally undoped, however, one expects an electron concentration of about $10^{8}$~cm$^{-2}$ due to a residual fluorine ions in the MBE chamber which serve as donors for ZnSe and Zn$_{0.85}$Mg$_{0.15}\text{Se}$.

Sample-3 (zq1038) is also a single QW structure ZnSe/Zn$_{0.89}$Mg$_{0.11}$S$_{0.18}$Se$_{0.82}$ with \textit{n}-type modulation doping. It has an 8\,nm-thick ZnSe QW, which is separated from the surface by a 50~nm-thick Zn$_{0.89}$Mg$_{0.11}$S$_{0.18}$Se$_{0.82}$ barrier. The doping layer (chlorine donors) of 3~nm thickness is separated from the QW by a 10~nm-thick spacer, and is located between the QW and the surface. Despite a nominal \textit{n}-type doping the charge redistribution leads to the presence of resident holes with a density of about $n_h=1\times10^{10}$~cm$^{-2}$. This is confirmed by the time-resolved pump-probe measurements~\cite{Zhu14} and by means of magneto-optical spectroscopy~\cite{Astakhov2002}.

Sample-4 (\#11376) contains ten layers of (In,Ga)As/GaAs QDs, separated by 100\,nm thick GaAs barriers. The QD density in each layer is about $1\times10^{10}$~cm$^{-2}$. The sample is thermally annealed for 30\,s at $960^\circ$C. Being nominally undoped, the sample contains fractions of charged QDs. Due to residual \textit{p}-type doping from carbon impurities in the MBE chamber we find a majority of QDs with hole charging. Further details on the sample characterization are given in Refs.~\cite{Crooker2010,Varwig2012,PhysRevB.98.121304}.

\subsection{Experimental techniques}\label{subsec:techniqes}

We use the pump-probe technique to study the electron and hole spin dynamics by time-resolved Kerr or Faraday rotation (KR or FR, respectively)~\cite{Awschalom_Spintronics,Kikkawa98,yugova12}. The circularly polarized pump pulses of 1.5~ps duration (spectral width of about 1\,meV) generated by a mode-locked Ti:Sapphire laser operating at a repetition frequency of 75.7~MHz (repetition period $T_{\text{R}}=13.2$~ns) are used for excitation. In the case of samples 2 and 3, the energy of the laser is doubled using a beta-barium borate crystal.

To avoid any dynamic nuclear polarization, the helicity of pump pulses is modulated between $\sigma^+$ and $\sigma^-$ polarizations by an electro-optical modulator (EOM) at the frequency $f_{\text{m}}$ ($f_{\text{m}}T_R\ll 1$), so that on average the samples are equally exposed to left and right circularly polarized pump pulses. The induced electron or hole spin coherence is measured by a linearly-polarized probe pulses of the same photon energy as the pump pulses (degenerate pump-probe scheme). The excitation energy $E$ and the pump power $P_{\rm pump}$ are given for each sample in the corresponding figure captions. The probe power is about one order of magnitude smaller in each case. The signal, proportional to the rotation angle, is measured by a balanced photoreceiver, connected to a lock-in amplifier. We use a double-modulation scheme of registration to reduce the effects of the scattered pump. Therefore, the probe beam is additionally modulated by the second EOM, and demodulation of the signal by the lock-in is done at a difference frequency of pump and probe modulation frequencies. Measurements for samples 2 and 3 were performed in the reflection geometry, while samples 1 and 4 are measured in the transmission geometry.

The samples are placed in a vector magnet system consisting of three superconducting split-coils oriented orthogonally to each other~\cite{Zhukov2012}. It allows us to measure the signal at the various magnetic field orientations from the Faraday geometry ($B_{\parallel}$) to the Voigt geometry ($B_{\perp}$) without changing of the optical alignment and conditions of signal detection. The measurements are performed at low temperatures in the range from $T=1.8$\,K up to 6\,K.

For our experiments, we have used three different variations of the pump-probe method:

(a) \textit{The time-resolved regime:} dependence of the Kerr rotation angle upon the time delay between the pump and probe pulses is measured with the magnetic field applied in the Voigt geometry ($B_{\perp}$). The Larmor precession of the optically oriented spins around the magnetic field results in a periodic signal with decreasing amplitude. Using this method one can determine the $g$ factor of the carriers and the inhomogeneous spin dephasing time $T_2^*$ in the case of $T_2^*<T_R$.

(b) \textit{The resonant spin amplification (RSA) regime} is used when the spin dephasing time $T_2^*$ is comparable or bigger than $T_R$. In this case the rotation angle is determined in dependence of $B_{\perp}$ at a fixed small negative time delay between pump and probe pulses (in our case $-50$\,ps)~\cite{Awschalom_Spintronics,Kikkawa98,yugova12}. Depending on the magnetic field the spin polarization is modulated, reaching maximum at the commensurability conditions of the Larmor frequency with the laser repetition period. Through the experimental part of the paper we consider the Hanle curve being identical to the RSA peak centered around zero magnetic field, the so called zero RSA peak.

(c) \textit{The polarization recovery curve (PRC) regime:} the electron spin polarization is detected at a small negative time delay in dependence of the magnetic field applied in the Faraday geometry ($B_{\parallel}$). In this case, the spin polarization is photogenerated along the magnetic field direction and does not exhibit Larmor precession on average. Still, the spin polarization will be reduced by the fluctuating nuclear fields if the external magnetic field is small compared to these fields.

\section{Experimental results}\label{sec:exp_results}

The whole set of shapes of the measured PRC signals can be divided into three groups:

\textit{The first group.} The classical PRC shape (V-shape): a constant signal amplitude (constant degree of carrier spins polarization) in the range of large longitudinal magnetic fields and a dip (a decrease of the amplitude) in the vicinity of the zero magnetic field. The PRC width is determined by the nuclear fields fluctuations~\cite{merkulov02} and coincides with the width of the zero RSA peak. Their amplitudes are related to each other as 2:1. The responsible spin relaxation mechanisms are discussed in Sec.~\ref{sec:basic}. Note that in some cases, e.g., in the presence of anisotropic hyperfine interaction, the PRC shape remains the same, but the width of the PRC and zero RSA peak and their amplitudes can be different (Sec.~\ref{sec:Anisotrop}, Fig.~\ref{fig:z}).

\textit{The second group.} The PRC has an additive shape: it consists of two pronounced V-shape components, typically with large difference in amplitudes and widths. However, the PRC has a dip at zero magnetic field and with increasing field the PRC amplitude reaches a constant value, as in the first case. Different components can belong to different groups of electrons, or to electrons and holes in the same structure.

\textit{The third group.} The PRC has an M-shape. With an increase of the magnetic field, the amplitude of the PRC initially increases, reaches the maximum value and then decreases (Sec.~\ref{sec:Pulsed excitation}).

\subsection{First group: basic type}\label{par:basic}

\begin{figure}[t]
\includegraphics[width=1.0\linewidth]{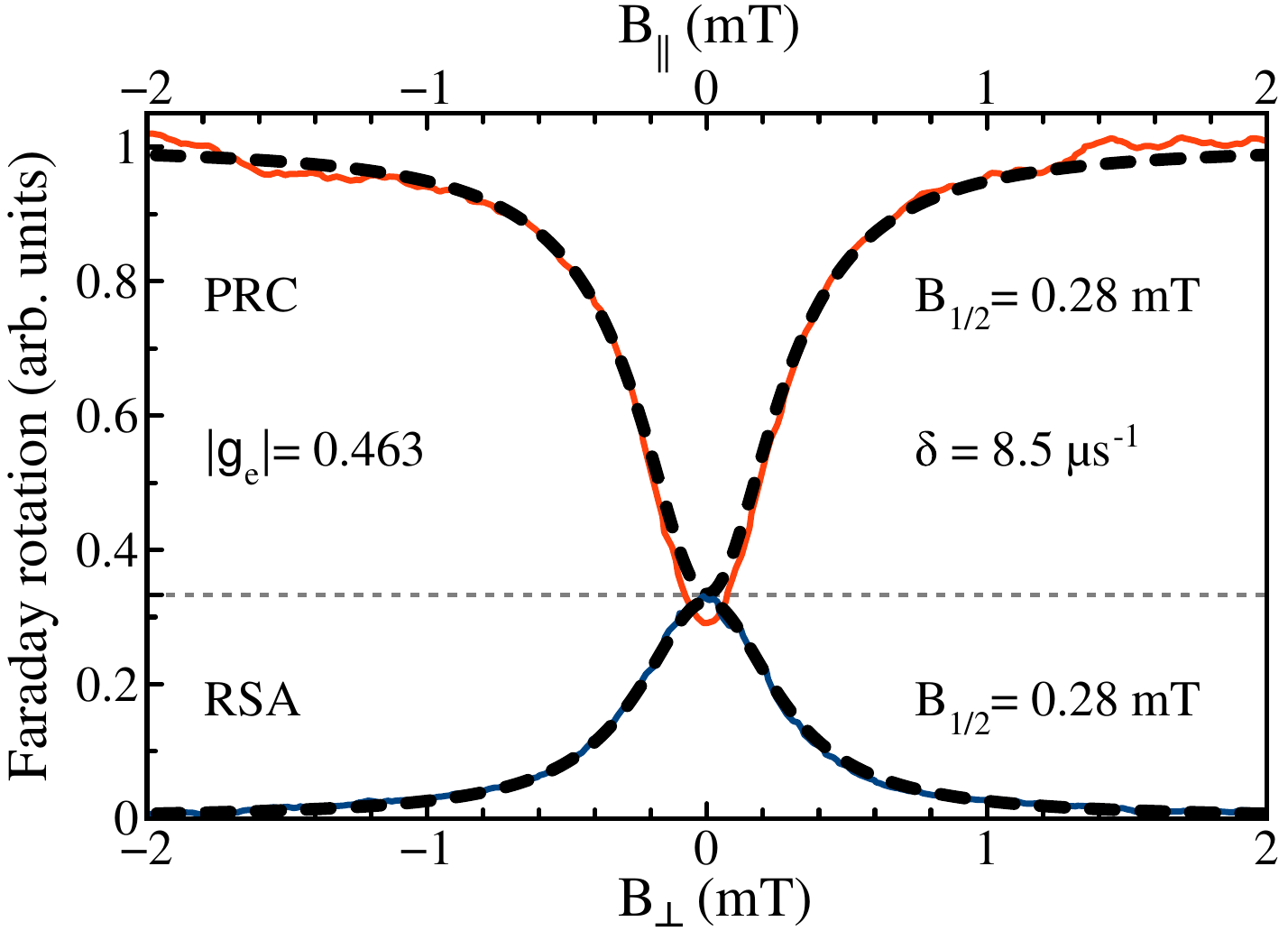}
\caption[]{$n$-doped bulk GaAs (sample-1). PRC (red) and RSA (blue) signals in dependence of magnetic fields $B_{\parallel}$ and $B_{\perp}$, respectively, at the same detection conditions.
They are fitted with Eq.~\eqref{eq:P_main} for PRC and Eq.~\eqref{eq:H_main} for zero RSA peak, as shown by the black dashed lines. Gray dashed line represents the basic 2 to 1 ratio of the amplitudes. $E=1.493$\,eV, $P_{\text{pump}}=1$\,W/cm$^{2}$, $f_{\text{m}} = 50$\,kHz, and $T=1.8$\,K.}
\label{fig:basic}
\end{figure}

In all following cases we use the time-resolved pump-probe measurements in constant magnetic field to define the $g$ factor of the corresponding charge carriers. In the sample-1 of $n$-doped bulk GaAs the measured value of the electron $g$ factor is $|g_e|=0.463$. We use it as a fixed parameter for the fitting. Figure~\ref{fig:basic} shows the normalized RSA and PRC at the same experimental conditions. The PRC is related to the top horizontal axis, while the RSA to the bottom one, both given in the same magnetic field range. Note that the chosen magnetic field range only shows the zero RSA peak, while the other RSA peaks are not seen.

The PRC in the sample-1 has the classical V-shape: electron spin polarization is increased with growing magnetic field and saturates in fields greater than 1\,mT. The measured half width at a half maximum (HWHM) of the PRC dip is $B_{1/2}=0.28$~mT. It is important to mention that it is equal to the HWHM of the zero RSA peak. From fitting of both curves, using Eq.~\eqref{eq:P_main} for the PRC and Eq.~\eqref{eq:H_main} for the zero RSA peak with the same set of parameters, we obtain $\delta=8.5$~$\mu$s$^{-1}$, see the black dashed lines in Fig.~\ref{fig:basic}. The ratio between the amplitudes of PRC and RSA signals at $B=0$ is close to the ratio of 2:1, see the dashed gray line in Fig.~\ref{fig:basic} at 1/3 in this scale~\cite{merkulov02}. These features of the PRC and RSA allow us to conclude that their shapes in weak magnetic fields are determined by the spin relaxation driven by the fluctuating nuclear magnetic fields. Therefore, they can be described in the framework of the basic model represented in Sec.~\ref{sec:basic}.

\subsection{First group: effects of nuclear spin correlation time}\label{par:nucl_corr}

\begin{figure}[t]
\includegraphics[width=1.0\linewidth]{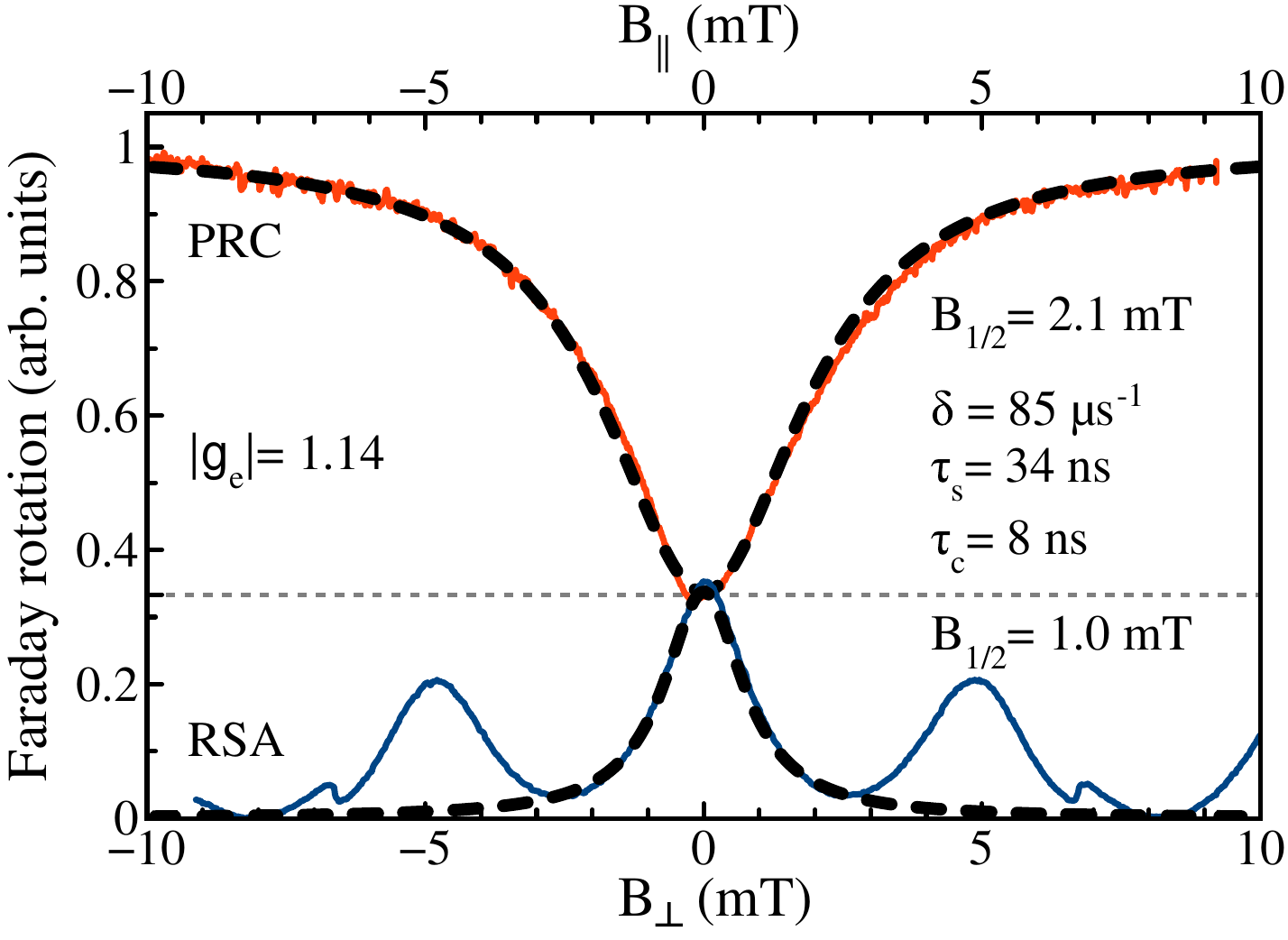}
\caption[] {\textit{n}-type ZnSe/Zn$_{0.85}$Mg$_{0.15}\text{Se}$ single QW (sample-2). PRC (red) and RSA (blue) signals detected at the same experimental conditions. Black dashed lines are the fits using the Eq.~\ref{eq:short_tau}. $E=2.8037$~eV, $P_{\text{pump}}=0.35$~W/cm$^{2}$, $f_{\text{m}} = 50$~kHz, and $T=1.8$~K.}
\label{fig:nucl_corr}
\end{figure}

As the second example in the first group we demonstrate results for the \textit{n}-type ZnSe/Zn$_{0.85}$Mg$_{0.15}\text{Se}$ single QW (sample-2). The dependence of KR signal on the time delay between pump and probe in a transverse magnetic field at resonant trion excitation of $E=2.8037$\,eV (not shown here) leads to $|g_e|=1.14$. The normalized PRC and RSA are shown in Fig.~\ref{fig:nucl_corr}. The PRC has a dip in vicinity of zero magnetic field and the PRC amplitude saturates in longitudinal magnetic fields exceeding 10\,mT. The HWHM of the dip is $B_{1/2}=2.1$~mT. The RSA demonstrates the classical behaviour of a periodic signal in increasing magnetic field. Due to a larger $g$ factor relative to the previous case, the RSA peaks are shifted closer to each other, which makes it possible to see several of them. For us only the zero RSA peak is of interest, which has a HWHM of $B_{1/2}=1.0$~mT, that is two times smaller compared to the PRC width. The ratio of the PRC and RSA amplitudes is close to 2:1~\cite{merkulov02}, which allows us to attribute this case to the first group of the basic type. However, to explain the difference in HWHM, we need to take into account the effect of the nuclear spin correlation time (Section~\ref{sec:Nuclear}). The situation corresponds to the case of the short correlations with $\tau_c \ll 1/\delta$. The fits by the black dashed lines in Fig.~\ref{fig:nucl_corr} support this assumption. We use Eq.~\eqref{eq:short_tau} with the set of parameters $\delta=85$\,$\mu$s$^{-1}$, $\tau_s=34$\,ns and $\tau_c=8$\,ns and obtain $\tau_c < 1/\delta = 11.8$\,ns $< \tau_s$. The short correlation time may be related with the electron hopping between the donors.

\subsection{Second group: additive spin components}\label{par:additive}

\begin{figure}[t]
\includegraphics[width=1.0\linewidth]{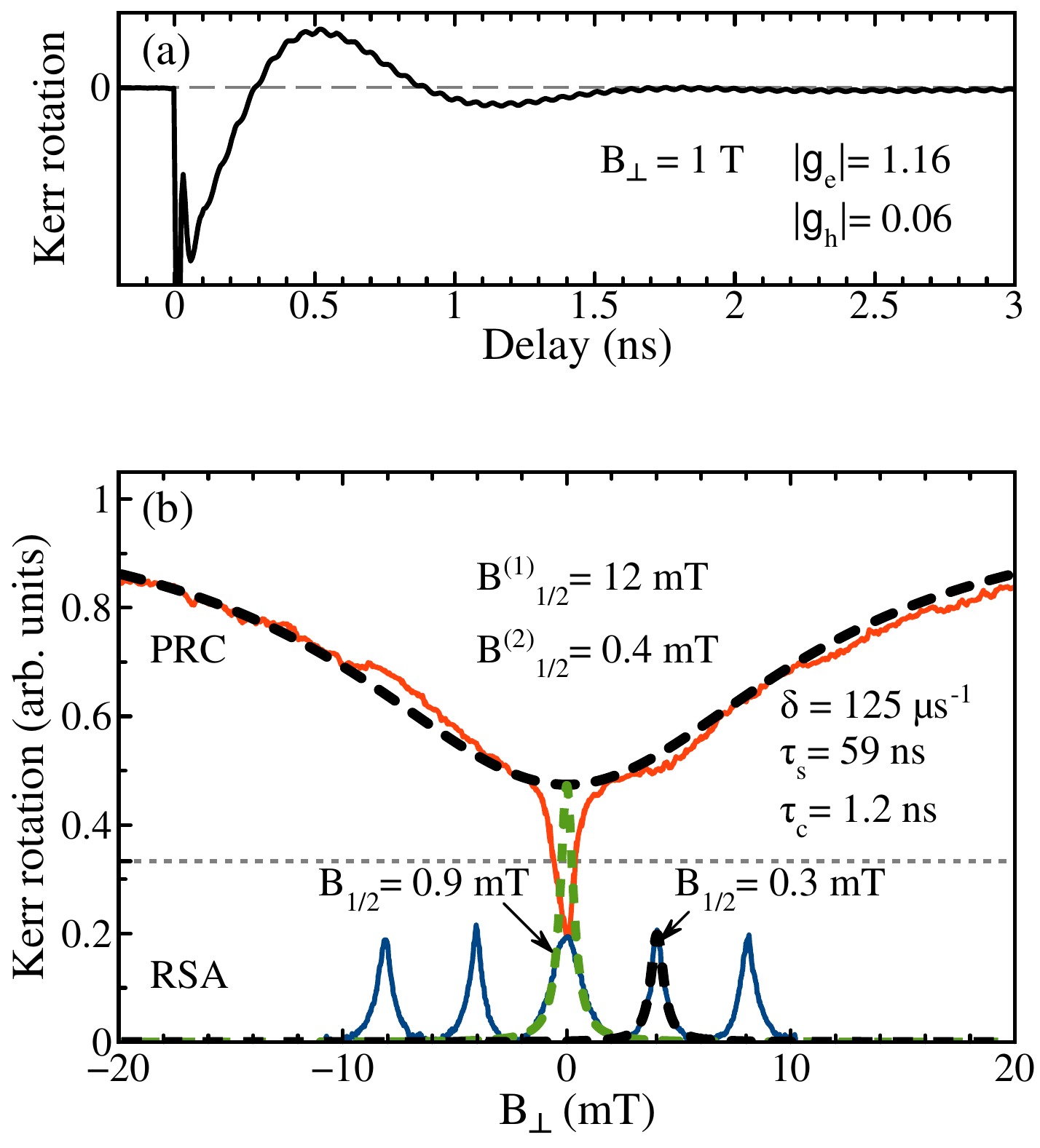}
\caption[] {ZnSe/Zn$_{0.89}$Mg$_{0.11}$S$_{0.18}$Se$_{0.82}$ single QW with resident holes (sample-3). (a) KR signal demonstrating two oscillating components at $B_{\perp}=1$\,T and corresponding $g$ factors. (b) PRC (red) and RSA (blue) signals with fittings (black dashed lines) using Eq.~\eqref{eq:short_tau}. The green dashed line shows the unmodified fit for the RSA peak. The gray dashed line at 1/3 is the reference of the basic case. $E=2.812$\,eV, $P_{\rm pump}=0.35$~W/cm$^{2}$, $f_\mathrm{m} = 100$\,kHz, and $T=1.8$\,K.}\label{fig:additive}
\end{figure}

Figure~\ref{fig:additive}(a) shows the dynamics of the KR signal for the ZnSe/Zn$_{0.89}$Mg$_{0.11}$S$_{0.18}$Se$_{0.82}$ single QW with resident holes (sample-3). It evidences the electron and hole spin precession, i.e. the presence of resident electrons and holes in this structure. The pump is resonant with the trion state ($E=2.812$~eV). The signal has two types of oscillations, a slow one with a frequency corresponding to $g$ factor of holes $|g_{h,\perp}|=0.06$ and a fast one with $g$ factor of electrons $|g_{e,\perp}|=1.16$, which are clearly seen at a transverse magnetic field of $B_{\perp}=1$\,T.

Let us first discuss the PRC signal, shown by red line in Fig.~\ref{fig:additive}(b). It is composed of two peaks: the broad one with $B^{(1)}_{1/2}=12$\,mT and the narrow one with $B^{(2)}_{1/2}=0.4$\,mT. The broad peak corresponds to the electrons, as one requires a much stronger external field for stabilization of the electron spin polarization due to the strong electron-nuclear interaction. Holes, on contrary, have a reduced hyperfine interaction with nuclei (and comparable longitudinal $g$ factor), which leads to the much narrower PRC width.

The RSA signal shown in Fig.~\ref{fig:additive}(b) show at five periodic peaks with the zero peak HWHM of 0.9\,mT. Other peaks have smaller width, e.g. of 0.3\,mT for the $\pm$1 peaks at about $\pm$4\,mT. We can relate this to the fact that the holes have a much smaller in-plane $g$ factor than electrons ($|g_{h,\perp}| = 0.06$ vs. $|g_{e,\perp}|= 1.16$). This leads to the difference in the RSA peaks separation for the both types of carriers. The hole RSA peaks would have a separation of about 90\,mT, while the electrons of about 4\,mT. Also, due to larger $g$-factor dispersion the holes have faster spin dephasing resulting in decreasing of the hole spin amplitude with increasing magnetic field. Therefore, in the sample-3, where the resident electrons and holes coexist, the zero RSA peak is contributed by both carriers, while the next peaks have only electron contribution. This allows us to analyze separately the electron and hole spin dynamics. For the fitting of the electron component in PRC and RSA by Eq.~\eqref{eq:short_tau} with one set of parameters we use RSA peak at 4\,mT and reduce the amplitude of the fit for the RSA peak in the Hanle part of Eq.~\eqref{eq:short_tau} by an additional factor of 2.4 to match the amplitude of the RSA to the PRC at $B=0$\,mT. The green dashed line in Fig.~\ref{fig:additive}(b) demonstrates the result of this fit with non-shifted and non-reduced peak for the electronic part of the zero RSA peak. The black dashed lines are the fits with the parameters $\delta=125$\,$\mu$s$^{-1}$, $\tau_s=59$\,ns and $\tau_c=1.2$\,ns. Here $1/\delta = 8$\,ns and, therefore, we are in the regime of $\tau_c \ll 1/\delta \ll \tau_s$.

To summarize, we have demonstrated that the PRC and RSA signals are influenced by the short nuclear spin correlation for the electron part and by the additive contribution of holes. The last fact complicates the analyzes, as the electron and hole contributions intermix with each other. The possibility to use the higher field RSA peaks shows the advantage of the pump-probe time-resolved technique compared to the typical Hanle measurements under continuous-wave excitation. It allows us to separate clearly the electron and hole contributions and to obtain spin parameters.

\subsection{Third group: M-shaped PRC}\label{par:M}
\label{sec:M-shape}

In all previous cases the strong carrier localization leads to the fact that the hyperfine interaction of electron spins with nuclei spin fluctuations is the main mechanism of spin relaxation in weak magnetic fields. Increase of the longitudinal magnetic field leads to an effective decoupling of the electron spins from the nuclear spins. As a result, a larger number of spins line up along the $z$ axis, and since we measure the $z$-spin component, the signal amplitude increases. All previous examples demonstrate the saturation of the electron spin polarization with an increasing $B_{\parallel}$. A fundamentally different PRC shape is observed in $p$-doped (In,Ga)As/GaAs quantum dots (sample-4), where the PRC amplitude decreases with a further increase of the field.

\begin{figure}[t]
\includegraphics[width=1.0\linewidth]{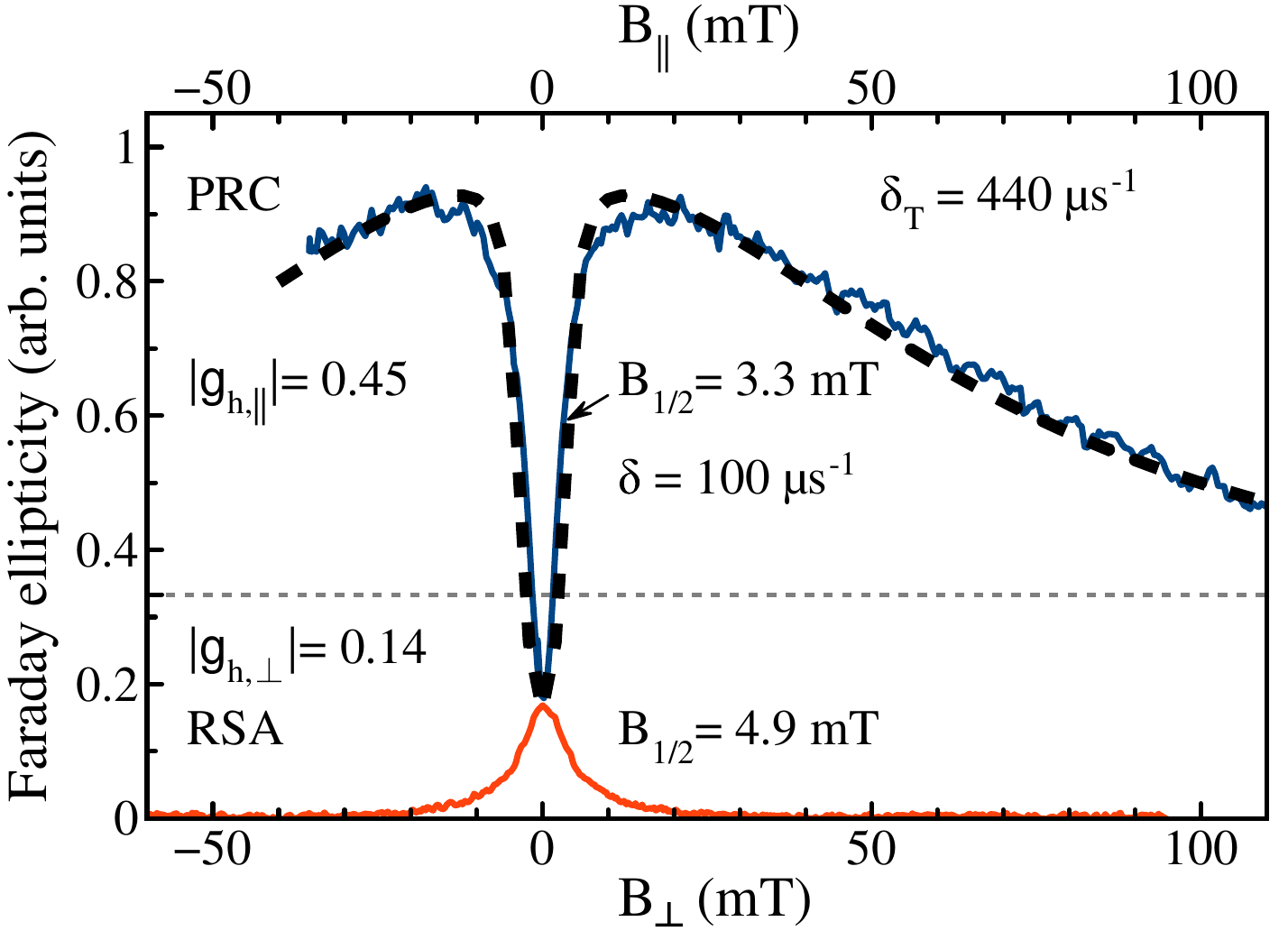}
\caption[] {$p$-doped (In,Ga)As/GaAs QDs (sample-4). PRC (blue) and RSA (red) signals at the same experimental conditions. Black dashed line is a fit using Eq.~\eqref{eq:S0P}~\cite{PhysRevB.98.121304}. $E=1.392$\,eV, $P_{\text{pump}}=0.35$~W/cm$^{2}$, $f_\mathrm{m}=25$\,kHz, and $T=1.8$\,K.}
\label{fig:pQDs}
\end{figure}

Figure~\ref{fig:pQDs} demonstrates the PRC for the sample-4 measured in ellipticity configuration (spin induced circular polarization of the transmitted light) as a function of the longitudinal magnetic field $B_{\parallel}$, the top axis. The observed M-shape of the PRC is determined by the spin dynamics of the resident hole in the ground state and the positive trion. The central narrow peak with $B_{1/2}=3.3$\,mT is driven by the interaction of the resident hole with the nuclear spins, while the much broader peak with decreasing amplitude at higher fields is driven by the interaction of the unpaired electron spin of the positive trion with the nuclear spins and the trion recombination~\cite{PhysRevB.98.125306}, as discussed in Sec.~\ref{subsec:M-form}. This interpretation is confirmed by a good agreement between the experimental data and the calculation (black dashed line in Fig.~\ref{fig:pQDs}) using Eq.~\eqref{eq:S0P} with $\delta = 100$\,$\mu$s$^{-1}$ for the resident holes and $\delta_T = 440$\,$\mu$s$^{-1}$ for the electron in the positive trion, for the whole set of parameters see Ref.~\cite{PhysRevB.98.121304}.

The RSA signal in the $p$-doped (In,Ga)As/GaAs quantum dots has only one peak centered around zero field. Higher field RSA peaks do not appear under these conditions due to several factors. First, the transverse hole $g$ factor is small, which results in larger separation of the RSA peaks in magnetic field scale. Second, the holes have a large spread of $g$ factors, in our case $\Delta g= 0.04$. This induces strong decrease of the RSA amplitude with increasing magnetic field. Furthermore, as demonstrated in Ref.~\cite{Varwig2012}, under specific excitation condition, the RSA peaks can switch to a constant non-zero amplitude without dependence on magnetic field, which is related to the mode-locking effect~\cite{PhysRevB.98.155318}, see also Sec.~\ref{subsec:pulsed_excitation} and Fig.~\ref{fig:cont_pulses}.

Additional effect, which should be taken into account is the influence of the anisotropic hyperfine interaction, as discussed in Sec.~\ref{sec:Anisotrop}. We show in Ref.~\cite{PhysRevB.98.121304} that in the $p$-doped (In,Ga)As/GaAs quantum dots (sample-4) the anisotropy parameter $\lambda=5$~\cite{PhysRevB.101.115302}, which should lead to the spin amplitude of about 0.75 at $B=0$\,mT, see Eq.~\eqref{eq:f}. However, this amplitude (in our case 0.17) and the widths of the RSA and PRC are further strongly influenced by the effects of the $g$-factor anisotropy ($|g_{h,\parallel}|=0.45$ \textit{vs.} $|g_{h,\perp}|=0.14$) and the pump power, see Sec.~\ref{subsec:pulsed_excitation} and  Fig.~\ref{fig:cont_pulses}. So, one can expect that due to a large anisotropy of the hole $g$ factor, the width of the PRC should be much smaller than the RSA one. On the other hand, the widths relation (together with the amplitudes) is additionally influenced by the pump power, where the PRC width gets strongly increased relative to the RSA curve, and the amplitude gets strongly reduced, for the higher pumping. Therefore, for a full understanding of the influence of different contributions, in this case power dependence would be helpful. We restrict ourself to the demonstration of the shape of the curves and leave the analyzes of the power dependence for the future investigations.

\section{Conclusion}\label{sec:concl}

To conclude, we analyzed theoretically the spin polarization recovery and Hanle effects accounting for the anisotropy of the hyperfine interaction, presence of two types of the charge carriers, nuclear spin dynamics and pulsed spin excitation. We also discussed the manifestations of the quantum Zeno and anti-Zeno effects in the pump-probe experiments as well as the resonant spin amplification in the Faraday geometry.
The experimental demonstrations were serving as examples to highlight the presence of the discussed effects. We show that the experimental signals are usually affected by several mechanisms. Using the methods of PRC and RSA with varying magnetic fields and the time-resolved measurements turn out to be an efficient way to study depolarizing effects of the nuclear spins on the electron and hole spins.

\acknowledgments

We are grateful to M.~M. Glazov, I.~A. Yugova, and P. Schering for valuable discussions. This work was partially financially supported by the RF President Grant No. MK-1576.2019.2 and the Foundation for the Advancement of Theoretical Physics and Mathematics ``BASIS''. We acknowledge partial financial support by the Deutsche Forschungsgemeinschaft in the frame of the International Collaborative Research Center TRR 160 (Project A1) and the Russian Foundation for Basic Research (Grants No. 19-52-12038 and 20-32-70048).

%\bibliography{PRC_bibliorgaphy}


\end{document}